\documentclass[a4paper,oneside,12pt]{myclass}

\usepackage{amsmath}    
\usepackage{amsfonts}   
\usepackage{euscript}   
\usepackage{feynmp}     
\usepackage{graphics}   
\usepackage{ifthen}     
\usepackage{slashed}    

\def\chaptername{Chapter}  
\def\appendixname{Appendix}
\def\bibname{References}   
\def\contentsname{Contents}
\newdimen\Left          
\newdimen\Right         
\textheight=\paperheight
\advance\textheight by-2\topmargin
\advance\textheight by-2\headheight
\advance\textheight by-2\headsep
\oddsidemargin=\Left
\evensidemargin=\Right
\textwidth=\paperwidth
\advance\textwidth by-\Left
\advance\textwidth by-\Right

\let\goth=\mathfrak              
\let\script=\mathcal             

\def\frak#1{\goth{#1}}           

\def\Hil{\script{H}}

\def\lie#1{\frak{#1}}             
\def\LA{\script{L}}                

\DeclareRobustCommand\openone{\leavevmode\hbox{\small1\normalsize\kern-.33em1}}
\def\vek#1{{\boldsymbol{\bf #1}}}
\def\he#1{#1^{\dagger}}         
\def\tr#1{#1^{\mathrm T}}                 
\def\bra#1{\langle#1\vert}
\def\ket#1{\vert#1\rangle}      
\def\braket#1#2{\langle#1\vert#2\rangle}

\def\de{\partial}
\def\vp{\varphi}

\def\imag{{\rm i}}              
\def\pubcite#1{[\expandafter\uppercase\expandafter{\romannumeral#1}]}

\def\abs#1{\lvert#1\rvert}                   

\def\dthree{\mathop{{\rm d}^3\!}\nolimits}
\def\dfour{\mathop{{\rm d}^4\!}\nolimits}    

\DeclareMathOperator{\re}{Re}                
\DeclareMathOperator{\im}{Im}                
\DeclareMathOperator{\Tr}{Tr}

\def\=={=\discretionary{}{\hbox{$=$}}{}}
\def\++{+\discretionary{}{\hbox{$+$}}{}}

\def\<<{<\discretionary{}{\hbox{$<$}}{}}
\def\>>{>\discretionary{}{\hbox{$>$}}{}}

\newlength{\captionwidth}
\makeatletter
\let\captionsize\footnotesize
\newsavebox{\tempbox}
\renewcommand{\@makecaption}[2]{
  \captionsize
  \sbox{\tempbox}{#1: #2}
  \ifthenelse{\lengthtest{\wd\tempbox > \linewidth}}
    {#1: #2\par}
    {\begin{center}#1: #2\end{center}}
  \normalsize}

\newcommand{\mojec@ption}[2]{%
  \captionsize
  \par
  \sbox{\tempbox}{#1: #2}
  \ifthenelse{\lengthtest{\wd\tempbox > \linewidth}}
    {\begin{center}\parbox[t]{\captionwidth}{#1: #2}\end{center}}
    {\begin{center}#1: #2\end{center}}
  \normalsize
}%
\let\@makecaption\mojec@ption
\makeatother

\begin{document}
\begin{fmffile}{fig}

\frontmatter



\vbox to\textheight{%
\begin{center}
\scalebox{1}{\includegraphics{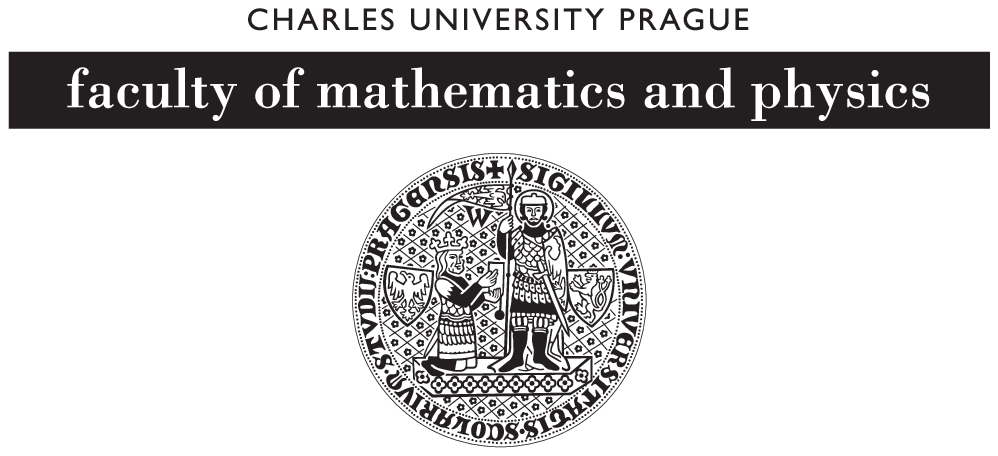}}
\vfill
\Huge\sffamily
Spontaneous symmetry breaking\\
in strong and electroweak interactions\\[3cm]
\Large\sffamily
Tom\'a\v{s} Brauner\\[1cm]
Nuclear Physics Institute\\
Academy of Sciences of the Czech Republic\\[5cm]
\large\sffamily
Thesis advisor: Ing. Ji\v{r}\'{\i} Ho\v{s}ek, CSc., NPI AS CR
\vfill
\end{center}}
\newpage


\section*{Acknowledgments}
I am indebted to my PhD advisor Ji\v{r}\'{\i} Ho\v{s}ek for his guidance
through the last four years and for sharing with me his insight into physics
during hours and hours of countless discussions. I am also grateful to
Professor Ji\v{r}\'{\i} Ho\v{r}ej\v{s}\'{\i} for his advice during the early
stages of my PhD study and for his continuous support.

I am obliged to Michael Buballa for his kind hospitality during my visit at the
Institute for Nuclear Physics, Technical University Darmstadt, and for the
amount of time he devoted to discussions with me.

I am thankful to Ji\v{r}\'{\i} Novotn\'{y} for enlightening discussions, and to
Ji\v{r}\'{\i} Ho\v{s}ek and Ji\v{r}\'{\i} Adam for the careful reading of the
manuscript of the thesis and several improving remarks and suggestions. I would
also like to acknowledge my fellow graduate students Petr Bene\v{s} and Adam
Smetana for the continuing collaboration.

The presented work was supported by the GACR grants No. 202/02/0847,
202/05/H003, and by the ASCR research plans No. K1048102, AV0Z10480505

My last and very personal word of thanks is reserved for \v{S}t\v{e}p\'anka:
Thank you for all your love and encouragement, without which this thesis and my
whole life wouldn't be what they are.

\section*{Declaration of originality}
This dissertation contains the results of research conducted at the Nuclear
Physics Institute of the Academy of Sciences of the Czech Republic in
\v{R}e\v{z}, in the period between fall 2002 and spring 2006. With the
exception of the introductory Chapter 2, and unless an explicit reference is
given, it is based on the published papers whose copies are attached at the end
of the thesis. I declare that the presented results are original. In some
cases, they have been achieved in collaboration as indicated by the authorship
of the papers.

\vskip2ex \hbox to \textwidth{Prague, April 3, 2006\hfill Tom\'a\v{s} Brauner}


{\parskip1ex
\tableofcontents}

\mainmatter
\setcounter{page}{4}


\chapter{Introduction}
The principle of spontaneous symmetry breaking underlies much of our current
understanding of the world around us. Although it has been introduced and
developed in full generality in particle physics, its applications also cover
a~large part of condensed matter physics, including such fascinating phenomena as
superconductivity, superfluidity, and Bose--Einstein condensation.

Ever since the very birth of science, philosophers, and later physicists,
admired the beauty of the laws of nature, one of their most appealing features
always being the \emph{symmetry}. Indeed, it was symmetry considerations that
lead Einstein to the creation of his theory of gravity, the general relativity,
and it is symmetry that is the basic building block of the modern theories of
the other fundamental interactions as well as all attempts to reconcile them
with Einstein's theory.

Symmetry is not only aesthetic, it is also practical. It provides an invaluable
guide to constructing physical theories and once applied, imposes severe
constraints on their structure. This philosophy has, in particular, lead to the
development of methods that allow us to exploit the symmetry content of the
system even if we actually cannot solve the equations of motion. The theory of
groups and their representations was first applied in quantum mechanics to the
problem of atomic and molecular spectra, and later in quantum field theory,
starting from the quark model and current algebra and evolving to the
contemporary gauge theories of strong and electroweak interactions, and the
modern concept of effective field theory.

There are many physical systems that, at first sight, display asymmetric
behavior, yet there is a~reasonable hope that they are described by symmetric
equations of motion. Such a~belief may be based, for instance, on the existence
of a~normal, symmetric phase, like in the case of superconductors and
superfluids. Another nice example was provided by the historical development of
the standard model of electroweak interactions. By the sixties, it was known
that the only renormalizable quantum field theories including vector bosons
were of the Yang--Mills type. It was, however, not clear how to marry the
non-Abelian gauge invariance of the Yang--Mills theory with the requirement
enforced by experiment, that the vector bosons be massive.

All these issues are resolved by the ingenious concept of a~\emph{spontaneously
broken symmetry}. The actual behavior of the physical system is determined by
the solution of the equations of motion, which may violate the symmetry even
though the action itself is symmetric. The internal beauty of the theory is
thus preserved and, moreover, one is able to describe simultaneously the normal
phase and the symmetry-breaking one. Just choose the solution which is
energetically more favorable under the specified external conditions.

This thesis presents a~modest contribution to the physics of spontaneous
symmetry breaking within the standard framework for the strong and electroweak
interactions and slightly beyond. The core of the thesis is formed by the
research papers whose copies are attached at the end. Throughout the text,
these articles are referred to by capital roman numbers in square brackets,
while the work of others is quoted by arabic numbers. The calculations
performed in the published papers are not repeated. We merely summarize the
results and provide a~guide for reading these articles and, to some extent,
their complement.

The thesis is a~collection of works on diverse topics, ranging from dynamical
electroweak symmetry breaking to color superconductivity of dense quark matter
and Goldstone boson counting in dense relativistic systems. Rather than giving
an exhaustive review of each of them, we try to keep clear the unifying concept
of spontaneous symmetry breaking and emphasize the similarity of methods used
to describe such vastly different phenomena.

Of course, such a~text cannot (and is not aimed to) be self-contained, and the
bibliography cannot cover all original literature as well. In most cases, only
those sources are quoted that were directly used in the course of writing. For
sake of completeness we quote several review papers where the original
references can also be found. The less experienced reader, e.g. a~student or
a~non-expert in the field, is provided with a~couple of references to lecture
notes on the topics covered.

The thesis is organized as follows. The next chapter contains an introduction
to the physics of spontaneous symmetry breaking. We try to be as general as
possible to cover both relativistic and nonrelativistic systems. The following
three chapters are devoted to the three topics investigated during the PhD
study. Chapter \ref{Chap:GBcounting} elaborates on the general problem of the
counting of Goldstone bosons, in particular in relativistic systems at finite
density. The electroweak interactions are considered in Chapter \ref{Chap:EWSB}
and an alternative way of dynamical electroweak symmetry breaking is suggested.
Finally, in Chapter \ref{Chap:QCD} we study dense matter consisting of quarks
of a~single flavor and propose a~novel mechanism for quark pairing, leading to
an unconventional color-superconducting phase. After the summary and concluding
remarks, the full list of author's publications as well as other references are
given. The reprints of the research papers published in peer-reviewed journals,
forming an essential and inseparable part of the thesis, are attached at the
end.


\chapter{Spontaneous symmetry breaking}
\label{Chap:SSB} In this chapter we review the basic properties of
spontaneously broken symmetries. First we discuss the general features, from
both the physical and the mathematical point of view. To illustrate the rather
subtle technical issues associated with the implementation of the broken
symmetry on the Hilbert space of states, a simple example is worked out in some
detail -- the Heisenberg ferromagnet.

After the general introduction we turn our attention to the methods of
description of spontaneously broken symmetries. We start with a~short
discussion of the model-independent approach of the effective field theory, and
then recall two particular models that we take up in the following chapters --
the linear sigma model and the Nambu--Jona-Lasinio model.

An extensive review of the physics of spontaneous symmetry breaking is given in
Ref. \cite{Guralnik:1968gh}. A~pedagogical introduction with emphasis on the
effective-field-theory description of Goldstone bosons may be found in the
lecture notes
\cite{Burgess:1998ku,Kaplan:2005es,Manohar:1996cq,Scherer:2002tk}.

\section{General features}
We shall be concerned with spontaneously broken \emph{continuous internal}
symmetries, that one meets in physics most often. The reason for such
a~restriction is twofold. First, this is exactly the sort of symmetries we
shall deal with in the particular applications to the strong and electroweak
interactions. Second, on the general ground, spontaneous breaking of discrete
symmetries does not give rise to the most interesting existence of Goldstone
bosons, while spacetime symmetries are more subtle, see Ref. \cite{Low:2001bw}.

As already noted in the Introduction, a~symmetry is said to be spontaneously
broken, if it is respected by the dynamical equations of motion (or,
equivalently, the action functional), but is violated by their particular
solution.\footnote{For a nice introductory account as well as several classical
examples see Refs. \cite{O'Raifeartaigh:1998sf,Straumann:1998yz}.} In quantum
theory we use, however, operators and their expectation values rather than
solutions to the classical equations of motion. Since virtually all information
about a~quantum system may be obtained with the knowledge of its ground state,
it is only necessary to define spontaneous breaking of a~symmetry in the ground
state or, the vacuum \cite{Coleman:1966co}.

\subsection{Realization of broken symmetry}
\label{Sec:Realization_broken_symmetry} Consider the group of symmetry
transformations generated by the charge $Q$. If the symmetry were a~true,
unbroken one, it would be realized on the Hilbert space of states by a~set of
unitary operators. In such a~case, their existence is guaranteed by the Wigner
theorem \cite{Weinberg:1995v1} and we speak of the \emph{Wigner--Weyl
realization} of the symmetry. The vacuum is assumed to be a~discrete,
\emph{nondegenerate} eigenstate of the Hamiltonian. Consequently, it bears
a~one-dimensional representation of the symmetry group, and therefore also is
an eigenstate of the charge $Q$. The excited states are organized into
multiplets of the symmetry, which may be higher-dimensional provided the
symmetry group is non-Abelian.

By this heuristic argument we have arrived at the definition of a~spontaneously
broken symmetry: \emph{A symmetry is said to be spontaneously broken if the
ground state is not an eigenstate of its generator Q.} A~very clean physical
example is provided by the ferromagnet. Below the Curie temperature, the
electron spins align to produce spontaneous magnetization. While the
Hamiltonian of the~ferromagnet is invariant under the $\mathrm{SU(2)}$ group of
spin rotations (not to be mixed up with spatial rotations -- see Section
\ref{Sec:Heis_ferro} for more details), this alignment clearly breaks all
rotations except those about the direction of the magnetization.

Note that as a~necessary condition for symmetry breaking it is usual to demand
just that the generator $Q$ does not annihilate the vacuum. Such a~criterion,
however, does not rule out the possibility that the ground state is an
eigenstate of $Q$ with nonzero eigenvalue. On the other hand, the vacuum charge
can always be set to zero by a~convenient shift of the charge operator.

A~distinguishing feature of broken symmetry is that the vacuum is infinitely
degenerate. In the case of the ferromagnet, the degeneracy corresponds to the
choice of the direction of the magnetization. In general, the ground states are
labeled by the values of a~symmetry-breaking order parameter. Formally, the
various ground states are connected by the broken-symmetry transformations.

With this intuitive picture in mind a~natural question arises, whether
a~physical system actually chooses as its ground state one of those with
a~definite value of the order parameter, or their superposition. To find the
answer, we go to finite volume and switch on a~weak external perturbation (such
as a~magnetic field). The degeneracy is now lifted and there is a~unique state
with the lowest energy. This mechanism is called \emph{vacuum alignment}.

After we perform the infinite volume limit and let the perturbation go to zero
(in this order), we obtain the appropriate ground state. In order for this
argument to be consistent, however, the resulting set of physically acceptable
vacua should not depend on the choice of perturbation. Indeed, it follows from
the general principles of causality and cluster decomposition that there is
a~basis in the space of states with the lowest energy such that all observables
become diagonal operators in the infinite volume limit \cite{Weinberg:1996v2}.

We have thus come to the conclusion that the correct ground state is one in
which the order parameter has a~definite value. The superpositions of such
states do not survive the infinite volume limit and therefore are not physical.
Moreover, transitions between individual vacua are not possible. This means
that rather than being a~set of competing ground states within a~single Hilbert
space, each of them constitutes a~basis of a~Hilbert space of its own, all
bearing inequivalent representations of the broken symmetry. This is called the
\emph{Nambu--Goldstone realization} of the symmetry.

To summarize, when a~symmetry is spontaneously broken, the vacuum is infinitely
degenerate. The individual ground states are labeled by the values of an order
parameter. In the infinite volume limit they give rise to physically
inequivalent representations of the broken symmetry. Transitions between
different spaces are only possible upon switching on an external perturbation.
This lifts the degeneracy and by varying it smoothly, one can adiabatically
change the order parameter.

This procedure can again be exemplified on the case of the ferromagnet. To
change the direction of the magnetization, one first imposes an external
magnetic field in the original direction of the magnetization. The magnetic
field is next rotated, driving the magnetization to the desired direction, and
afterwards switched off.

The issue of inequivalent realizations of the broken symmetry has rather subtle
mathematical consequences \cite{Guralnik:1968gh}, which we now shortly discuss
and later, in Section \ref{Sec:Heis_ferro}, demonstrate explicitly on the case
of the ferromagnet. As already mentioned, the Hilbert spaces with different
values of the order parameter are connected by broken-symmetry transformations.
The reason why they are called inequivalent is that these broken-symmetry
transformations are not represented by unitary operators. They merely provide
formal mappings between the various Hilbert spaces. By the same token, the
generator $Q$ is not a~well defined operator in the infinite volume limit. What
is well defined is just its commutators with other operators, which generate
infinitesimal symmetry transformations.

Since the broken symmetry is not realized by unitary operators, it is also not
manifested in the multiplet structure of the spectrum. This is determined by
the unbroken part of the symmetry group. Let us, however, stress the fact that
the broken symmetry is by no means similar to an approximate, but spontaneously
unbroken one. Even though it does not generate multiplets in the spectrum, it
still yields \emph{exact constraints} which must be satisfied by, e.g., the
Green's functions of the theory.

\subsection{Goldstone theorem}
\label{Sec:Goldstone_theorem} One of the most striking consequences of
spontaneous symmetry breaking is the existence of soft modes in the spectrum,
ensured by the celebrated Goldstone theorem
\cite{Goldstone:1961eq,Goldstone:1962es}. In its most general setting
applicable to relativistic as well as nonrelativistic theories, it can be
formulated as follows: \emph{If a~symmetry is spontaneously broken, there must
be an excitation mode in the spectrum of the theory whose energy vanishes in
the limit of zero momentum.} In the context of relativistic field theory this,
of course, means that the so-called \emph{Goldstone boson} is a~massless
particle.

Several remarks to the Goldstone theorem are in order. First, in the general
case it does not tell us how many Goldstone modes there are. Anyone who learned
field theory in the framework of particle physics knows that in
Lorentz-invariant theories, the number of Goldstone bosons is equal to the
number of broken-symmetry generators \cite{Weinberg:1996v2}. In the
nonrelativistic case, however, the situation is more complex and there is in
fact no completely general counting rule that would tell us the exact number of
the Goldstone modes. This issue will be discussed in much more detail in
Chapter \ref{Chap:GBcounting}.

Second, there are technical assumptions which, in some physically interesting
cases, may be avoided, thus invalidating the conclusions of the Goldstone
theorem. A~sufficient condition for the theorem to hold is the causality which
is inherent in relativistic field theories. The nonrelativistic case is, again,
more complicated. In general, the Goldstone theorem applies if the potential
involved in the problem decreases fast enough towards the spatial infinity. An
example in which this condition is not satisfied is provided by the
superconductors where the long-range Coulomb interaction lifts the energy of
the low-momentum would-be Goldstone mode, producing a~nonzero gap
\cite{Fetter:1971fw}.

\begin{figure}
\begin{center}
\input{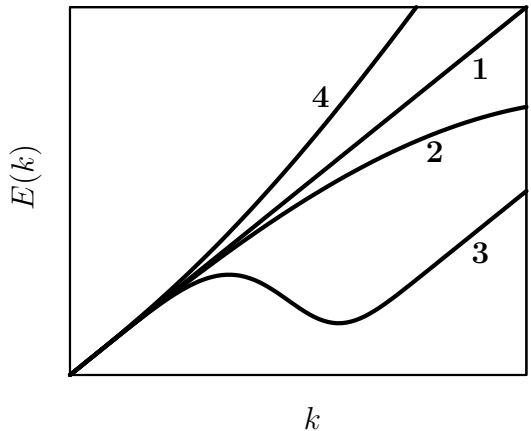}
\end{center}
\caption{Dispersion relations of the Goldstone bosons in four physically
distinct systems, conveniently normalized to have the same slope at the origin.
1. The Goldstone boson in a~relativistic field theory. 2. The acoustic phonon
in a~solid. 3. The phonon-roton excitation in the superfluid helium. 4. The
phonon in the relativistic linear sigma model at finite chemical potential (see
Chapter \ref{Chap:GBcounting}).} \label{Fig:GB_UV_disp_rel}
\end{figure}

Third, the Goldstone theorem gives us information about the low-momentum
behavior of the dispersion relation of the Goldstone boson. In the absence of
other gapless excitations, the long-distance physics is governed by the
Goldstone bosons and can be conveniently described by an effective field
theory. This does not tell us, however, anything about the high-energy
properties of the Goldstone bosons. At high energy, the dispersion relation of
the Goldstone mode is strongly affected by the details of the short-distance
physics. It is thus not as simple and universal as the low-energy limit, but at
the same time not uninteresting, as documented by Fig.
\ref{Fig:GB_UV_disp_rel}.

Let us now briefly recall the proof of the Goldstone theorem. The starting
assumption is the existence of a~conserved current, $j^{\mu}(x)$. From its
temporal component, the charge operator generating the symmetry is formed,
$$
Q(t)=\int\dthree\vek x\,j^0(\vek x,t).
$$
The domain of integration is not indicated in this expression. The charge
operator itself is well defined only in finite volume, but as long as its
commutators with other operators are considered, the integration may be safely
extended to the whole space \cite{Guralnik:1968gh}.

Now the broken-symmetry assumption about the ground state $\ket0$ is that
a~(possibly composite) operator $\Phi$ exists such that
\begin{equation}
\bra0[Q,\Phi]\ket0\neq0.
\label{Goldstone_commutator1}
\end{equation}
Note that this immediately yields our previous intuitive definition of broken
symmetry: The vacuum cannot be an eigenstate of the charge Q. This vacuum
expectation value is precisely what we called an order parameter before.

Inserting a~complete set of intermediate states into Eq.
\eqref{Goldstone_commutator1} and assuming the translation invariance of the
vacuum, one arrives at the representation
\begin{equation}
\bra0[Q,\Phi]\ket0=\sum_n(2\pi)^3\delta(\vek k_n)\left[e^{-\imag E(\vek k_n)t}
\bra0j^0(0)\ket n\bra n\Phi\ket0-e^{\imag E(\vek k_n)t}\bra0\Phi\ket n\bra
nj^0(0)\ket0\right]. \label{Goldstone_commutator2}
\end{equation}

Using the current conservation one can show that the Goldstone commutator in
Eq. \eqref{Goldstone_commutator1} is time-independent provided the surface term
which comes from the integral,
$$
\int\dthree\vek x\,[\nabla\cdot\vek j,\Phi],
$$
vanishes. This is the central technical assumption which underlies the
requirements of causality or fast decrease of the potential mentioned above.

Once this condition is satisfied, the time independence of the Goldstone
commutator forces the right-hand side of Eq. \eqref{Goldstone_commutator2} to
be time-independent as well. This is, however, not possible unless there is
a~mode in the spectrum such that $\lim_{\vek k\to0}E(\vek k)=0$, which is the
desired Goldstone boson.

\section{Toy example: Heisenberg ferromagnet}
\label{Sec:Heis_ferro} The general statements about spontaneous symmetry
breaking will now be demonstrated on the Heisenberg ferromagnet. Consider
a~cubic lattice with a~spin-$\frac12$ particle at each site. The dynamics of
the spins is governed by the Hamiltonian
\begin{equation}
H=-J\sum_{\text{pairs}}\vek s_i\cdot\vek s_j,
\label{Heisenberg_Hamiltonian}
\end{equation}
which is invariant under simultaneous rotations of all the spins, that form the
group $\mathrm{SU(2)}$.

For simplicity we choose the \emph{nearest-neighbor interaction} so that the
sum in Eq. \eqref{Heisenberg_Hamiltonian} runs only over the pairs of
neighboring sites. The coupling constant $J$ is assumed positive so that the
interaction favors parallel alignment of the spins. In finite volume we shall
take up the periodic boundary condition in order to preserve the (discrete)
translation invariance of the Hamiltonian \eqref{Heisenberg_Hamiltonian}.

\subsection{Ground state}
The scalar product of two neighboring spin operators may be simplified to
$$
\vek s_i\cdot\vek s_j=\tfrac12\left[(\vek s_i+\vek s_j)^2-(\vek s_i^2+\vek
s_j^2)\right]=\tfrac12(\vek s_i+\vek s_j)^2-\tfrac34.
$$
It is now clear that the state with the lowest energy will be one in which all
pairs of spins will be arranged to have total spin one. The scalar product
$\vek s_i\cdot\vek s_j$ then reduces to $\tfrac14$. In a~three-dimensional
cubic lattice with $N$ sites in total, there are altogether $3N$ such pairs so
that the ground-state energy of the ferromagnet is
$$
E_0=-\tfrac34NJ.
$$

As we learned in the course of our general discussion of broken symmetries, the
ground state is infinitely degenerate. The individual states may be labeled by
the direction of the magnetization, a~unit vector $\vek n$. All spins are
aligned to point in this direction, which means that the ground state vector
$\ket{\Omega(\vek n)}$ is a~direct product of one-particle states, the
eigenvectors of the operators $\vek n\cdot\vek s_i$ with eigenvalue one half,
$$
\ket{\Omega(\vek n)}=\prod_{i=1}^N\ket{i,\vek n},\quad\text{where}\quad(\vek
n\cdot\vek s_i)\ket{i,\vek n}=\tfrac12\ket{i,\vek n}.
$$

The one-particle states may be expressed explicitly in terms of the two
spherical angles $\theta,\vp$ in the basis of eigenstates of the third
component of the spin operator,
\begin{equation}
\ket{\vek n}=\left(
\begin{array}{c}
\cos\tfrac\theta2\\
e^{\imag\vp}\sin\tfrac\theta2
\end{array}\right).
\label{spin_eigenvector}
\end{equation}

The two vectors $\ket{i,\vek n}$ and $\ket{i,-\vek n}$ form an orthonormal
basis of the one-particle Hilbert space $\Hil_i$. The products of these vectors
then constitute a~basis of the full Hilbert space of the ferromagnet,
$\Hil=\bigotimes_{i=1}^N\Hil_i$.

In finite volume $N$, states with all possible directions $\vek n$ can be
accommodated within a~single Hilbert space. Two one-particle bases $\{\ket{\vek
n_1},\ket{-\vek n_1}\}$ and $\{\ket{\vek n_2},\ket{-\vek n_2}\}$ are, as usual,
connected by the unitary transformation corresponding to the rotation that
brings the vector $\vek n_1$ to the vector $\vek n_2$. Likewise, the two
corresponding product bases of the full Hilbert space $\Hil$ are connected by
the induced unitary rotation on this product space.

Let us now calculate the scalar product of the ground states assigned to two
directions $\vek n_1$ and $\vek n_2$. By exploiting the rotational invariance
of the system, we may rotate one of the vectors, say $\vek n_1$, to the
$z$-axis. The explicit expression for the eigenvectors \eqref{spin_eigenvector}
then yields $\braket{\vek n_1}{\vek n_2}=\cos\tfrac{\theta_{\vek n_1,\vek
n_2}}2$, where $\theta_{\vek n_1,\vek n_2}$ is the angle between the two unit
vectors.

The scalar product of the two ground-state vectors is then given by
$$
\braket{\Omega(\vek n_1)}{\Omega(\vek n_2)}=\left(\cos\tfrac{\theta_{\vek n_1,\vek
n_2}}2\right)^N
$$
and it apparently goes to zero as $N\to\infty$ unless $\vek n_1$ and $\vek n_2$
are (anti)parallel.

Using a~slightly different formalism we shall now construct the whole Hilbert
space $\Hil(\vek n)$ above the ground state $\ket{\Omega(\vek n)}$ and show
that, in fact, any two vectors, one from $\Hil(\vek n_1)$ and the other from
$\Hil(\vek n_2)$, are orthogonal in the limit $N\to\infty$.

Recall that the two-dimensional space of spin $\tfrac12$ may be viewed as the
Fock space of the fermionic oscillator. One defines an annihilation operator
$a(\vek n)$ and a~creation operator $\he a(\vek n)$ so that
$$
a(\vek n)\ket{\vek n}=0\quad\text{and}\quad\{a(\vek n),\he a(\vek n)\}=1.
$$
These are actually nothing else than the lowering and raising operators
familiar from the theory of angular momentum. In addition to the identities
above, they satisfy
$$
[a(\vek n),\he a(\vek n)]=2\vek n\cdot\vek s,\quad\text{so that}\quad
\vek n\cdot\vek s=-\he a(\vek n)a(\vek n)+\tfrac12.
$$
When $\vek n=(0,0,1)$, these operators are just $a=s_x+\imag s_y$, $\he
a=s_x-\imag s_y$, and in the general case they can be found explicitly by the
appropriate unitary rotation.

The Hilbert space $\Hil(\vek n)$ is set up as a~Fock space above the vacuum
$\ket{\Omega(\vek n)}$. In the ground state all spins point in the direction
$\vek n$, while the excited states are obtained by the action of the creation
operators $\he a_i(\vek n)$ that flip the spin at the $i$-th lattice site to
the opposite direction.\footnote{Note that, in this setting, annihilation and
creation operators at different lattice sites \emph{commute} rather than
anticommute as usual. The change of sign induced by the interchange of two
distinguishable fermions is, however, merely a~convention.} The basis of the
space $\Hil(\vek n)$ contains all vectors of the form $\he a_{i_1}(\vek n)\he
a_{i_2}(\vek n)\dotsb\ket{\Omega(\vek n)}$ where a~\emph{finite} number of
spins are flipped.

It is now obvious that in the infinite-volume limit, all basis vectors from the
space $\Hil(\vek n_1)$ are orthogonal to all basis vectors from the space
$\Hil(\vek n_2)$ that is, these two spaces are completely orthogonal.

To put it in yet another way, at finite $N$ any vector from the space
$\Hil(\vek n_1)$ may be expressed as a~linear combination of the basis vectors
of the space $\Hil(\vek n_2)$, and thus these two spaces may be identified.
This is, however, no longer true as $N\to\infty$, for the linear combination in
question then contains an infinite number of terms, and is divergent. There is
no other way out than treating the spaces $\Hil(\vek n_1)$ and $\Hil(\vek n_2)$
as distinct, orthogonal ones.

To summarize, in the limit $N\to\infty$ one has a~continuum of mutually
orthogonal separable Hilbert spaces $\Hil(\vek n)$ labeled by the direction of
the magnetization $\vek n$. In the absence of explicit symmetry breaking no
transition between different spaces is possible and one has to choose the
vector $\vek n$ once for all and work within the space $\Hil(\vek n)$.
Operators representing the observables are then constructed from the
annihilation and creation operators $a_i(\vek n)$ and $\he a_i(\vek n)$.

The symmetry transformations are formally generated by the operator of the
total spin, $\vek S=\sum_i\vek s_i$. It is now evident that those
transformations that change the direction of the magnetization $\vek n$, i.e.
the spontaneously broken ones, are not realized by unitary operators since they
do not operate on the Hilbert space $\Hil(\vek n)$. The only operator that does
is the projection of the total spin on the direction of the magnetization,
$\vek n\cdot\vek S$. This generates the unbroken subgroup. (Yet, this operator
is unbound for $N\to\infty$, but it can be normalized by dividing by $N$ to
yield the spin density, which is already finite.)

It is worth emphasizing, however, that physically all directions $\vek n$ are
equivalent. Measurable effects can only arise from the change of the direction
of $\vek n$.

\subsection{Goldstone boson}
\label{Sec:magnon} One may now ask where is the Goldstone boson associated with
the spontaneous breakdown of the $\mathrm{SU(2)}$ symmetry of the Hamiltonian
\eqref{Heisenberg_Hamiltonian}. In the general discussion of the Goldstone
theorem we assumed full translation invariance, while this lattice system has
only a~discrete one. Fortunately, this is not a~problem in the infinite-volume
limit, where there is still a~continuous momentum variable $\vek k$ to label
one-particle states. The only difference is that only a~finite domain of
momentum, the Brillouin zone, should be used. We shall therefore assume that
$-\pi\slash\ell\leq k_x,k_y,k_z\leq+\pi\slash\ell$, where $\ell$ is the lattice
spacing.

As we emphasized above, all directions of $\vek n$ are physically equivalent,
so we shall from now on set $\vek n=(0,0,1)$. The scalar product of two
neighboring spins may be rewritten in terms of the annihilation and creation
operators,
\begin{multline}
\vek s_i\cdot\vek s_j=\tfrac14(a_i+\he a_i)(a_j+\he a_j)-\tfrac14(a_i-\he
a_i)(a_j-\he a_j)+(-\he a_ia_i+\tfrac12)(-\he a_ja_j+\tfrac12)=\\
=-\tfrac12(\he a_i-\he a_j)(a_i-a_j)+\he a_ia_i\he a_ja_j+\tfrac14.
\label{Heis_Ham_second_quantization}
\end{multline}

Note that the Hamiltonian preserves the `particle number' that is, the number
of flipped spins generated by the operator $\sum_i\he a_ia_i$. This is of
course, up to irrelevant constants, nothing but the third component of the
total spin, which is not spontaneously broken and thus can be used to label
physical states. We shall restrict our attention to the `one-particle' space,
spanned on the basis $\ket i=\he a_i\ket{\Omega(\vek n)}$. The physical reason
behind this restriction is that the sought Goldstone boson turns out to be the
spin wave -- a~traveling perturbation induced by flipping a~single spin.

On the one-particle space, the second term on the right hand side of Eq.
\eqref{Heis_Ham_second_quantization} gives zero while the constant $\tfrac14$
may be dropped. The one-particle Hamiltonian thus reads
$$
H_{1\text P}=\frac J2\sum_{\text{pairs}}(\he a_i-\he a_j)(a_i-a_j),
$$
and acts on the basis states as\footnote{As the low-energy dynamics of the
Goldstone boson is isotropic, we work without lack of generality in one space
dimension. The index $i$ now refers to the linear ordering of the spin chain.}
\begin{equation}
H_{1\text P}\ket i=-\frac J2\bigl(\ket{i+1}-2\ket i+\ket{i-1}\bigr).
\label{spin_wave_ham}
\end{equation}

The discrete translation invariance is apparently not broken in the ground
state. That means that the stationary states are simultaneously the eigenstates
of the shift operator, $T:\ket i\to\ket{i+1}$. The eigenvalues of the shift
operator are of the form $e^{\imag k\ell}$. Eq. \eqref{spin_wave_ham} implies
that the one-particle Hamiltonian is diagonalized in the basis of eigenstates
of $T$. The corresponding energies are
\begin{equation}
E(k)=\frac J2\left(2-e^{\imag k\ell}-e^{-\imag
k\ell}\right)=2J\sin^2\frac{k\ell}2, \label{magnon_disp_rel}
\end{equation}
and in three dimensions we would analogously find $E(\vek
k)=2J\bigl(\sin^2\frac{k_x\ell}2+\sin^2\frac{k_y\ell}2+\sin^2\frac{k_z\ell}2\bigr)$.

We have thus found our Goldstone boson, in the case of the ferromagnet it is
called the \emph{magnon}. We stress the fact that we used no approximation, so
Eq. \eqref{magnon_disp_rel} is the exact dispersion relation of the magnon, and
the eigenstate $\sum_je^{\imag jk\ell}\ket j$ is the exact eigenstate of the
full Hamiltonian \eqref{Heisenberg_Hamiltonian}.

Note also that there is just one Goldstone mode even though two symmetry
generators, $S_x$ and $S_y$, are spontaneously broken. This may be intuitively
understood by acting with either broken generator on the vacuum
$\ket{\Omega(\vek n)}$. We find $S_x\ket{\Omega(\vek n)}=\tfrac12\sum_j\ket j$
and $S_y\ket{\Omega(\vek n)}=\tfrac\imag2\sum_j\ket j$ that is, both operators
create the same state, which formally corresponds to the zero-momentum magnon.
This fact appears to be tightly connected to the dispersion relation of the
magnon, which is quadratic at low momentum. The phenomenon is quite general and
its detailed discussion is deferred to Chapter \ref{Chap:GBcounting}.

Having found the exact dispersion relation, it is suitable to comment on the
issue of finite vs. infinite volume. Strictly speaking, there is no spontaneous
symmetry breaking in finite volume. All the effects such as the unitarily
inequivalent implementations of the symmetry and the existence of a~gapless
excitation appear only in the limit of infinite volume. Real physical systems
are, on the other hand, always finite-sized. They are, however, large enough
compared to the intrinsic microscopic scale (here the lattice spacing $\ell$)
of the theory so that the infinite-volume limit is both meaningful and
practical.

In particular, when the ferromagnet lattice is of finite size $N$, the periodic
boundary condition requires the momentum $k$ to be quantized, the minimum
nonzero value being $k_{\text{min}}\ell=2\pi/N$. The energy gap in the magnon
spectrum is then $E_{\text{min}}\approx 2\pi^2J/N^2$, which is small enough for
any macroscopic system to be to set to zero.

\section{Description of spontaneous symmetry breaking}
\label{Sec:DescriptionSSB} So far we have been discussing the very general
features of spontaneously broken symmetries. To investigate a~physical system
in more detail, one next has to fix the Lagrangian. Before going into
particular models we shall make an aside and mention the very important concept
of effective field theory.

The method of effective field theory relies on the fact that, in the absence of
other gapless excitations, the long-distance physics of a~spontaneously broken
symmetry is governed by the Goldstone bosons.\footnote{Quite generally, the
effective field theory approach may be applied whenever there are two or more
energy scales in the system which can be treated separately. It is thus not
special only to spontaneous symmetry breaking. This philosophy is emphasized in
the lecture notes by Kaplan \cite{Kaplan:2005es} and Manohar
\cite{Manohar:1996cq}.} One then constructs the most general effective
Lagrangian for the Goldstone degrees of freedom, compatible with the underlying
symmetry \cite{Weinberg:1996v2}.

The chief advantage of this approach is that it provides a~model-independent
description of the broken symmetry. The point is that by exploiting the
underlying symmetry, it essentially yields the most general parametrization of
the observables in terms of a~set of low-energy coupling constants.

From the physical point of view, a~disadvantage of effective field theory is
that it tells us nothing about the origin of symmetry breaking -- one simply
has to assume a~particular form of the symmetry-breaking pattern.

To show that the symmetry is broken at all and to specify the symmetry-breaking
pattern, one has to find an appropriate order parameter. It is therefore not
surprising that the issue of finding a~suitable order parameter is of key
importance, and considerable difficulty, for the description of spontaneous
symmetry breaking.

In the following, we recall two particular models of spontaneous symmetry
breaking. The operator whose vacuum expectation value provides the order
parameter is an elementary field in the first case, and a~composite object in
the second one. In both cases, an approximation is made such that the quantum
fluctuations of the order parameter are neglected.

\subsection{Linear sigma model}
\label{Sec:Linear_sigma_model} Perhaps the most popular and universal approach
to spontaneous symmetry breaking is to construct the Lagrangian so that it
already contains the order parameter. This is very much analogous to the
Ginzburg--Landau theory of second-order phase transitions. One introduces
a~scalar field\footnote{The order parameter has to be a~scalar unless one wants
to break the space-time symmetry \cite{Sannino:2002wp}.} and adjusts the
potential so that it has a~nontrivial minimum. The result is the paradigmatic
Mexican hat.

The great virtue of this method is that the order parameter is provided by the
vacuum expectation value of an elementary scalar field, which may be chosen
conveniently to achieve the desired symmetry-breaking pattern. As a~particular
example we shall now review the simplest model with Abelian symmetry.

Starting with a~pure scalar theory, we define the Lagrangian for a~complex
scalar field $\phi$ as
\begin{equation}
\LA_{\phi}=\de_{\mu}\he\phi\de^{\mu}\phi+M^2\he\phi\phi-\lambda(\he\phi\phi)^2.
\label{Abelian_sigma_Lagrangian}
\end{equation}
This Lagrangian is invariant under the phase transformations $\phi\to\phi
e^{\imag\theta}$ that form the Abelian group $\mathrm{U(1)}$. At tree level,
the ground state is determined by the minimum of the static part of the
Lagrangian, which is found at $\he\phi\phi=v^2/2=M^2/2\lambda$ so that the
symmetry is spontaneously broken. As explained in Section
\ref{Sec:Realization_broken_symmetry}, there is a~continuum of solutions to
this condition (distinguished by their complex phases) and the physical vacuum
may be chosen as any one of them, but not their superposition. This is the
reason why the following classical analysis actually works.

It is customary to choose the order parameter real and positive i.e., we set
$\langle\phi\rangle=v/\sqrt2$. The scalar field is next shifted to the minimum
and parametrized as $\phi=(v+H+\imag\pi)/\sqrt2$. Upon this substitution the
Lagrangian becomes
$$
\LA_{\phi}=\tfrac12(\de_{\mu}H)^2+\tfrac12(\de_{\mu}\pi)^2+\tfrac14M^2v^2-M^2H^2
-\lambda vH^3-\tfrac14\lambda H^4-\tfrac12\lambda H\pi^2-\tfrac14\lambda
H^2\pi^2-\tfrac14\lambda\pi^4.
$$
The first three terms represent the kinetic terms for $H$ and $\pi$ and minus
the vacuum energy density, respectively. There is also the mass term for $H$,
while the field $\pi$ is massless -- this is the Goldstone boson.

It is instructive to evaluate the $\mathrm{U(1)}$ Noether current in terms of
the new fields,
\begin{equation}
j^{\mu}=\imag(\he\phi\de^{\mu}\phi-\de^{\mu}\he\phi\phi)=
-v\de^{\mu}\pi+(\pi\de^{\mu}H-H\de^{\mu}\pi).
\label{current_pion_contribution}
\end{equation}
We can see that the Goldstone boson is annihilated by the broken-symmetry
current, as predicted by the Goldstone theorem. The corresponding matrix
element is given by $\bra0j^{\mu}(0)\ket{\pi(\vek k)}\propto vk^{\mu}$, the
constant of proportionality depending on the normalization of the one-particle
states.

In the standard model of electroweak interactions, the scalar field is in fact
added just for the purpose of breaking the gauge and global symmetries of the
fermion sector. The same may be done in our toy model. We start with a~free
massless Dirac field whose Lagrangian,
$\LA_{\psi}=\bar\psi\imag\slashed\de\psi$, is invariant under the
$\mathrm{U(1)_V}\times\mathrm{U(1)_A}$ chiral group. The mass term of the
fermion violates the axial part of the symmetry and thus can be introduced only
after this is broken.

To that end, we add the scalar field Lagrangian $\LA_{\phi}$ and an interaction
term $\LA_{\phi\psi}=y(\bar\psi_L\psi_R\phi+\bar\psi_R\psi_L\he\phi)$. The full
Lagrangian, $\LA=\LA_{\psi}+\LA_{\phi}+\LA_{\phi\psi}$, remains chirally invariant
provided the scalar $\phi$ is assigned a~proper axial charge. The nontrivial
minimum of the potential in Eq. \eqref{Abelian_sigma_Lagrangian} now breaks the
axial symmetry spontaneously and, upon the reparametrization of the scalar
field, the fermion acquires the mass $m=vy/\sqrt2$.

\subsection{Nambu--Jona-Lasinio model}
\label{Sec:NJL_model} In contrast to the phenomenological linear sigma model
stands the idea of dynamical spontaneous symmetry breaking. Here, one does not
introduce any artificial degrees of freedom in order to break the symmetry by
hand but rather tries to find a~symmetry-breaking solution to the quantum
equations of motion.

Physically, this is the most acceptable and ambitious approach. Unfortunately,
it is also much more difficult than the previous one. The reason is that one
often has to deal with strongly coupled theories and, moreover, the
calculations always have to be nonperturbative. As a~rule, it is usually simply
assumed that a~symmetry-breaking solution exists and after it is found, it is
checked to be energetically more favorable than the perturbative vacuum.

By this sort of a~variational argument, one is able to prove that the symmetric
perturbative vacuum is not the true ground state. On the other hand, it does
not follow that the found solution is, which might be a~problem in complex
systems where several qualitatively different candidates for the ground state
exist \cite{Rajagopal:2000wf}.

As an example, we shall briefly sketch the model for dynamical breaking of
chiral symmetry invented by Nambu and Jona-Lasinio
\cite{Nambu:1961tp,Nambu:1961fr,Klevansky:1992qe}. As the same model will be
used in Chapter \ref{Chap:QCD} to describe a~color superconductor
\cite{Buballa:2003qv}, we shall take up this opportunity to introduce the
mean-field approximation that we later employ.

The Lagrangian of the original Abelian NJL model reads
\begin{equation}
\LA=\bar\psi\imag\slashed\de\psi+G\left[(\bar\psi\psi)^2-(\bar\psi\gamma_5\psi)^2\right].
\label{NJL_Lagrangian}
\end{equation}
Its invariance under the Abelian chiral group
$\mathrm{U(1)_V}\times\mathrm{U(1)_A}$ is most easily seen when the interaction
is rewritten in terms of the chiral components of the Dirac field,
$\LA=\bar\psi\imag\slashed\de\psi+4G\abs{\bar\psi_R\psi_L}^2$.

Following the original method due to Nambu and Jona-Lasinio, we anticipate
spontaneous generation of the fermion mass by the interaction and split the
Lagrangian into the \emph{massive} free part and an interaction,
$\LA=\LA_{\text{free}}+\LA_{\text{int}}$, where
$$
\LA_{\text{free}}=\bar\psi(\imag\slashed\de-m)\psi,\quad
\LA_{\text{int}}=m\bar\psi\psi+G\left[(\bar\psi\psi)^2-(\bar\psi\gamma_5\psi)^2\right].
$$
At this stage already, we are making the choice of the ground state by
introducing the mass term and requiring that $m$ be real and positive. The
general parametrization of the mass term would be $\bar\psi(m_1+\imag
m_2\gamma_5)\psi$ with real $m_1,m_2$. The physical mass of the fermion would
then be $\sqrt{m_1^2+m_2^2}$.

The actual value of the mass $m$ is determined by the condition of
self-consistency, that it receives no one-loop radiative corrections. This
gives rise to the gap equation
\begin{equation}
1=8\imag G\int\frac{\dfour k}{(2\pi)^4}\frac1{k^2-m^2}.
\label{NJL_gap_equation}
\end{equation}

The same result may be obtained with a~method due to Hubbard and Stratonovich,
which keeps the symmetry of the Lagrangian manifest at all stages of the
calculation. One adds to the Lagrangian a~term
$-\abs{\phi-4G\bar\psi_R\psi_L}^2/4G$. In the path integral language, this
amounts to an additional Gaussian integration over $\phi$ that merely
contributes an overall numerical factor. Eq. \eqref{NJL_Lagrangian} then
becomes
\begin{equation}
\LA=\bar\psi\imag\slashed\de\psi-\frac1{4G}(\phi_1^2+\phi_2^2)+
\bar\psi(\phi_1+\imag\phi_2\gamma_5)\psi,
\label{Hubbard_Lagrangian1}
\end{equation}
the $\phi_1,\phi_2$ being the real and imaginary parts of $\phi$, respectively.

The Lagrangian is now bilinear in the Dirac field so that this may be
integrated out, yielding an effective action for the scalar order parameter
$\phi$,
\begin{equation}
\mathcal S_{\text{eff}}=-\frac1{4G}\int\dfour x\,
(\phi_1^2+\phi_2^2)-\imag\log\det\left[\imag\slashed\de+
(\phi_1+\imag\phi_2\gamma_5)\right] \label{Hubbard_Lagrangian2}
\end{equation}
With this effective action one can evaluate the partition function, or the
thermodynamic potential, in the saddle-point approximation. This means that we
have to replace the dynamical field $\phi$ with a~constant determined as
a~solution to the stationary-point condition,
$$
\frac{\delta\mathcal S_{\text{eff}}}{\delta\phi_1}=\frac{\delta\mathcal
S_{\text{eff}}}{\delta\phi_2}=0.
$$

Looking back at Eq. \eqref{Hubbard_Lagrangian1} we see that the constant
\emph{mean field} $\phi$ yields precisely the effective mass of the fermion,
and the stationary-point condition,
$$
1=8\imag G\int\frac{\dfour k}{(2\pi)^4}\frac1{k^2-\he\phi\phi},
$$
is identical to the gap equation \eqref{NJL_gap_equation}.

In the Nambu--Jona-Lasinio model, the Goldstone boson required by the Goldstone
theorem is a bound state of the elementary fermions. In the simple case of the
Lagrangian \eqref{NJL_Lagrangian} it is a pseudoscalar and may be revealed as a
pole in the two-point Green's function of the composite operator
$\bar\psi\gamma_5\psi$ \cite{Nambu:1961tp}.


\chapter{Goldstone boson counting in nonrelativistic systems}
\label{Chap:GBcounting} This chapter is devoted to a~detailed discussion of the
issue raised in Section \ref{Sec:Goldstone_theorem}: How many Goldstone bosons
are there, given the pattern of spontaneous symmetry breaking? As already
mentioned, in Lorentz-invariant theories the situation is very simple: The
number of Goldstone bosons is equal to the number of the broken-symmetry
generators. In nonrelativistic systems, however, these two numbers may differ.

We have already met an example where this happens -- the ferromagnet.
Historically, this was perhaps the first case in which the `abnormal' number of
Goldstone bosons was reported, and it still remains the only textbook one.
Nevertheless, the same phenomenon has recently been studied in some
relativistic systems at finite density
\cite{Miransky:2001tw,Schaefer:2001bq,Blaschke:2004cs,Beraudo:2004zr} as well
as in the Bose--Einstein condensed atomic gases \cite{Ho:1998ho,Ohmi:1998om},
and it is therefore desirable to analyze the problem of the Goldstone boson
counting on a~general ground.

We start with a~review of the general counting rule by Nielsen and Chadha
\cite{Nielsen:1976hm} and some other partial results. The main body of this
chapter then consists of the discussion of the Goldstone boson counting in the
framework of the relativistic linear sigma model at finite chemical potential.
The presented results are based on the paper \pubcite3, where the details of
the calculations may be found.

\section{Review of known results}
\subsection{Nielsen--Chadha counting rule}
Following closely the treatment of Nielsen and Chadha \cite{Nielsen:1976hm}, we
consider a~continuous symmetry, some of whose generators, $Q_a$, are
spontaneously broken. The broken-symmetry assumption
\eqref{Goldstone_commutator1} now generalizes to
$$
\det\bra0[Q_a,\Phi_i]\ket0\neq0,\quad a,i=1,\dotsc,\text{\# of broken
generators}.
$$

In addition, it is assumed that the translation invariance is not entirely
broken and that for any two local operators $A(x)$ and $B(x)$ a~constant
$\tau>0$ exists such that
\begin{equation}
\abs{\bra0[A(\vek x,t),B(0)]\ket0}\to e^{-\tau\abs{\vek x}}
\quad\text{as}\quad\abs{\vek x}\to\infty.
\label{commutativity_condition}
\end{equation}

It is then asserted that there are two types of Goldstone bosons -- type-I, for
which the energy is proportional to an odd power of momentum, and type-II, for
which the energy is proportional to an even power of momentum in the
long-wavelength limit. \emph{The number of Goldstone bosons of the first type
plus twice the number of Goldstone bosons of the second type is always greater
or equal to the number of broken generators.}

The difference between the two types of Goldstone bosons is nicely demonstrated
on the contrast between the ferromagnet and the antiferromagnet. In the
ferromagnet, there is a~single Goldstone boson (the magnon). The
Nielsen--Chadha counting rule then enforces that it must be of type II and
indeed, its dispersion relation is quadratic at low momentum, see Section
\ref{Sec:magnon}. In the antiferromagnet, on the other hand, there are two
distinct magnons with different polarizations. Their dispersion relation is
linear.

Note that the result of Nielsen and Chadha does not restrict in any way the
power of momentum to which the energy is proportional. As far as the counting
of the Goldstone bosons is concerned, it only matters whether this power is an
odd or an even number. It seems, however, that there are in fact no systems of
physical interest where the power is greater than two.

It is also worthwhile to mention that the Nielsen--Chadha counting rule is
formulated as an \emph{inequality}, in most cases of physical interest this
inequality is, however, saturated. This happens not only for the ferromagnet
and the antiferromagnet. To the best of the author's knowledge, all exceptions
where a~sharp inequality occurs, happen at a~phase boundary of the theory
\cite{Schaefer:2001bq,Sannino:2001fd}. Later in this chapter we shall see
a~generic class of such exceptions: The phase transition to the Bose--Einstein
condensed phase of the theory, at which the phase velocity of the superfluid
phonon vanishes and the phonon thus becomes a~type-II Goldstone boson.

It is natural to ask what is the difference between the ferromagnet and the
antiferromagnet that causes such a~dramatic discrepancy in their behavior. The
answer lies in the nonzero net magnetization of the ferromagnet. In general, it
is nonzero vacuum expectation values of some of the charge operators that
distinguish the type-II Goldstone bosons from the type-I ones. At a~very
elementary level, one can say that nonzero charge densities break time reversal
invariance and thus allow for the presence of odd powers of energy in the
effective Lagrangian for the Goldstone bosons \cite{Burgess:1998ku}. The issue
of charge densities, however, deserves more attention because they are usually
easier to determine than the Goldstone boson dispersion relations.

\subsection{Other partial results}
As we have just shown, the issue of Goldstone boson counting is tightly
connected to densities of conserved charges. We thus deal with three distinct
features of spontaneously broken symmetries that are related to each other: The
Goldstone boson counting, the charge densities in the ground state, and the
dispersion relations of the Goldstone bosons.

The connection between the Goldstone boson counting and the dispersion
relations is enlightened by the Nielsen--Chadha counting rule. In general,
little is known about the direct relation of the Goldstone boson counting and
the charge densities. There is a~partial (in fact, only negative) result of
Schaefer et al. \cite{Schaefer:2001bq} who proved that \emph{the number of
Goldstone bosons is usual i.e., equal to the number of broken generators,
provided the commutators of all pairs of broken generators have zero density in
the ground state.}

A~necessary condition for an abnormal number of Goldstone bosons is thus
a~nonvanishing vacuum expectation value of a~commutator of two broken
generators. The value of this result is that it shows that the pattern of
symmetry breaking must involve the non-Abelian structure of the symmetry group.
For instance, the Goldstone boson counting is usual in all
color-superconducting phases of QCD in which only the net baryon number density
is nonzero. The reason is that the baryon number corresponds to
a~$\mathrm{U(1)}$ factor of the global symmetry group and therefore does not
give rise to an order parameter for spontaneous symmetry breaking.

Intuitively, the necessity to modify the counting of the Goldstone bosons in
the presence of charge densities can be understood as follows \pubcite3. Assume
that the commutator of the charges $Q_a$ and $Q_b$ develops nonzero
ground-state expectation value. We may then in Eq.
\eqref{Goldstone_commutator2} set $Q=Q_a$ and take the charge density
$j^0_b(x)$ in place of the interpolating field for the Goldstone boson, $\Phi$.
We find
\begin{equation}
\imag f_{abc}\bra0
j^0_c(0)\ket0=\bra0[Q_a,j^0_b(x)]\ket0=2\imag\im\sum_n(2\pi)^3\delta(\vek
k_n)\bra0j^0_a(0)\ket n\bra nj^0_b(0)\ket0,
\label{charge_order_parameter}
\end{equation}
where $f_{abc}$ are the set of structure constants of the symmetry group. Two
points here deserve a~comment. First, it is again clear that a~non-Abelian
symmetry group is needed. Only then may the vacuum charge density be treated as
an order parameter for spontaneous symmetry breaking. Second, it follows from
the right hand side of Eq. \eqref{charge_order_parameter} that a~single
Goldstone boson couples to two broken currents, $j^{\mu}_a$ and $j^{\mu}_b$. We
have already seen in Section \ref{Sec:Heis_ferro} that this happens in the case
of the ferromagnet. This suggests the way how the counting rule for the
Goldstone bosons should be modified once nonzero density of a~non-Abelian
charge is involved. Nevertheless, it still remains to turn this heuristic
argument into a~more rigorous derivation of the proper counting rule.

Finally, the connection between the charge densities and the Goldstone boson
dispersion relations was provided by the work of Leutwyler
\cite{Leutwyler:1994gf}. Leutwyler analyzed spontaneous symmetry breaking in
nonrelativistic translationally and rotationally invariant systems. He
determined the leading-order low-energy effective Lagrangian for the Goldstone
bosons as the most general solution to the Ward identities of the symmetry. His
results show that when a~non-Abelian generator develops nonzero ground-state
density, a~term with a~single time derivative appears in the effective
Lagrangian. The time reversal invariance is then broken and the leading-order
Lagrangian is of the Schr\"odinger type, resulting in the quadratic dispersion
relation of the Goldstone boson. It should perhaps be stressed that when this
happens, the effective Lagrangian is invariant with respect to the prescribed
symmetry only up to a~total derivative.

We shall now give a~simple argument, also due to Leutwyler, explaining how such
a~single-time-derivative term in the Lagrangian affects the Goldstone boson
counting. The effective Lagrangian is constructed on the coset space of the
broken symmetry. Consequently, the number of independent \emph{real fields}
appearing in the Lagrangian is always equal to the number of broken generators.

Now if the single-time-derivative term is absent in the Lagrangian, the
Goldstone boson dispersion relation is linear and comes, at tree level, in the
form $E^2\propto\vek k^2$. This equation has both positive and negative energy
solutions which may be combined into a~single real scalar field (similar to the
Klein--Gordon field). There is therefore a~one-to-one correspondence between
the Goldstone bosons and the fields in the Lagrangian.

On the other hand, if there is a~term with a~single time derivative in the
Lagrangian, the Goldstone boson dispersion relation is quadratic and appears as
$E\propto\vek k^2$. This equation has, of course, only positive energy
solutions, very much like the Schr\"odinger equation. As a~result, the type-II
Goldstone boson is to be described with a~\emph{complex field} or,
equivalently, with a~pair of real fields. This shows why the type-II Goldstone
bosons have to be counted twice, when comparing their number to the number of
broken generators.

Now and again, this intuitive picture easily accommodates only the Goldstone
bosons with linear or quadratic dispersion. The question of the existence of
Goldstone bosons with energy proportional to higher powers of momentum remains
open as well as the possibility of their description in terms of a~low-energy
effective Lagrangian. Note that to achieve the appropriate power of momentum in
the dispersion law, one would have to get rid of the standard bilinear kinetic
term in the Lagrangian, which would invalidate the conventional perturbation
expansion as well as the power-counting scheme.

\section{Linear sigma model at finite chemical potential}
The rest of this chapter is devoted to the study of a~particular class of
Lorentz-noninvariant systems -- relativistic theories at finite density. The
microscopic dynamics of such systems is Lorentz-invariant, Lorentz symmetry
being violated only at the macroscopic level, by medium effects. This suggests
that much more could be said about the patterns of symmetry breaking and
properties of the Goldstone bosons than the Nielsen--Chadha theorem does, by
exploiting the underlying Lorentz invariance.

In the following, we shall stay in the framework of the relativistic linear
sigma model and derive an exact correspondence between the Goldstone boson
counting, charge densities, and the Goldstone boson dispersion laws. The
discussion of the possible extension of the achieved results is postponed to
the Conclusions.

\subsection{$\mathrm{SU(2)}\times\mathrm{U(1)}$ invariant sigma model}
We start with a simple example: The linear sigma model with an
$\mathrm{SU(2)}\times\mathrm{U(1)}$ symmetry, which has been used as a~toy
model for kaon condensation in the Color-Flavor-Locked phase of QCD
\cite{Miransky:2001tw,Schaefer:2001bq}. All essential steps leading to the
final counting rule for the Goldstone bosons will be first demonstrated within
this model, then within a~more complicated one with an
$\mathrm{SU(3)}\times\mathrm{U(1)}$ symmetry, and afterwards generalized to the
sigma model with arbitrary symmetry.

The model is defined by the Lagrangian,
\begin{equation}
\LA=D_{\mu}\he\phi D^{\mu}\phi-M^2\he\phi\phi-\lambda(\he\phi\phi)^2,
\label{Lagrangian_kaon_cond}
\end{equation}
where the scalar $\phi$ is a~complex doublet. Nonzero density of the
$\mathrm{U(1)}$ charge is implemented in terms of the chemical potential $\mu$,
which enters the Lagrangian through the covariant derivative,
$D_0\phi=(\de_0-\imag\mu)\phi$.

In the absence of the chemical potential, the Lagrangian
\eqref{Lagrangian_kaon_cond} is invariant under the extended group
$\mathrm{SU(2)}\times\mathrm{SU(2)}\simeq\mathrm{SO(4)}$. The chemical
potential breaks it explicitly down to $\mathrm{SU(2)}\times\mathrm{U(1)}$. In
the context of the CFL phase with the kaon condensate, the $\mathrm{SU(2)}$
group corresponds to the isospin and the $\mathrm{U(1)}$ to the strangeness.
The field $\phi$ is just the (charged or neutral) kaon doublet.

The chemical potential contributes a~term $\mu^2\he\phi\phi$ to the static part
of the Lagrangian. When $\mu>M$, the perturbative vacuum $\phi=0$ becomes
unstable and a~new, nontrivial minimum appears -- the
$\mathrm{SU(2)}\times\mathrm{U(1)}$ symmetry is spontaneously broken down to
its $\mathrm{U(1)}$ subgroup. This is the relativistic Bose--Einstein
condensation.

To reveal the physical content of the model in the spontaneously broken phase,
we proceed in the standard manner i.e., calculate the minimum of the potential,
shift the scalar field, and expand the Lagrangian about the new ground state.
The scalar field is reparametrized as
$$
\phi=\frac1{\sqrt2}e^{\imag\pi_k\tau_k/v}\left(
\begin{array}{c}
0 \\ v+H
\end{array}\right),\quad\text{where}\quad
v^2=\frac{\mu^2-M^2}{\lambda},
$$
$\tau_k$ being the Pauli matrices. The three `pion' fields $\pi_k$ would, in
the absence of the chemical potential, correspond to the three Goldstone bosons
of the coset $[\mathrm{SU(2)}\times\mathrm{U(1)}]/\mathrm{U(1)}$.

The excitation spectrum is determined by the bilinear part of the Lagrangian,
\begin{equation}
\mathcal L_{\text{bilin}}=\frac12(\partial_{\mu}\pi_k)^2+
\frac12(\partial_{\mu}H)^2-v^2\lambda H^2
+\mu(\pi_1\partial_0\pi_2-\pi_2\partial_0\pi_1)
+\mu(H\partial_0\pi_3-\pi_3\partial_0H).
\label{bilin_Lagr}
\end{equation}
The presence of the chemical potential apparently leads to nontrivial mixing of
the fields which cannot be removed by a~global unitary transformation. To find
the dispersion laws of the four degrees of freedom, it is therefore more
appropriate to look for the poles of the propagators. It turns out
\cite{Miransky:2001tw,Schaefer:2001bq} that the mixing of $\pi_1$ and $\pi_2$
gives rise to \emph{one} Goldstone boson with the low-momentum dispersion law
$E(\vek k)=\vek k^2/2\mu$, while the other mode is gapped, $E(\vek k)
=2\mu+\mathcal{O}(\vek k^2)$. On the other hand, the sector $(\pi_3,H)$
produces one gapless excitation with $E(\vek k)
=\sqrt{\frac{\mu^2-M^2}{3\mu^2-M^2}}\abs{\vek k}+\mathcal{O}(\abs{\vek k}^3)$,
and a~massive radial mode with a~gap $\sqrt{3\mu^2-M^2}$.

In conclusion, there are two Goldstone bosons, one with a linear dispersion law
(the phonon) and one with a quadratic dispersion law. This is in accord with
the Nielsen--Chadha counting rule since the vacuum expectation value
$\langle\phi\rangle$ carries nonzero isospin. To see in more detail how this
fact affects the structure of the bilinear Lagrangian \eqref{bilin_Lagr}, note
that
$$
\mu(\pi_1\partial_0\pi_2-\pi_2\partial\pi_1)=-\frac\mu{v^2}
\pi_k\partial_0\pi_l\im\bigl\langle[\tau_k,\tau_l]\bigr\rangle.
$$

In this form it is obvious how the nonzero density of the commutator of two
broken charges \eqref{charge_order_parameter} enters the Lagrangian and thus
gives rise to the existence of a~single type-II Goldstone boson instead of two
type-I ones.

To understand more deeply the nature of the type-II Goldstone boson, we shall
now investigate the corresponding plane-wave solution of the classical equation
of motion. Note first that the unbroken $\mathrm{U(1)}$ group is generated by
the matrix $\tfrac12(1+\tau_3)$. In order to keep this $\mathrm{U(1)}$ symmetry
manifest, we combine $\pi_1$ and $\pi_2$ into one complex field,
$\psi=\tfrac1{\sqrt2}(\pi_2+\imag\pi_1)$. In fact, $\psi$ is nothing but the
upper component of the original doublet $\phi$, expanded to first order in
$\pi$.

As far as the quadratic Goldstone boson is concerned, we may drop the fields
$\pi_3$ and $H$ and rewrite the Lagrangian \eqref{bilin_Lagr} in terms of
$\psi$,
$$
\LA_{\psi}=2\imag\mu\he\psi\de_0\psi+\de_{\mu}\he\psi\de^{\mu}\psi.
$$
The field $\psi$ annihilates the type-II Goldstone and the corresponding
classical plane-wave solution is given by $\psi=\psi_0e^{-\imag k\cdot x}$,
with the exact (tree-level) dispersion relation
$$
E(\vek k)=\sqrt{\vek k^2+\mu^2}-\mu.
$$

The $\mathrm{SU(2)}\times\mathrm{U(1)}$ symmetry gives rise to four conserved
currents which, in terms of the doublet $\phi$, read
$$
j_k^{\mu}=-2\im\phi^{\dagger}\tau_k\partial^{\mu}\phi+
2\mu\delta^{\mu0}\phi^{\dagger}\tau_k\phi,\quad
j^{\mu}=-2\im\phi^{\dagger}\partial^{\mu}\phi+
2\mu\delta^{\mu0}\phi^{\dagger}\phi.
$$
For the quadratic Goldstone plane wave we find
$$
j_1^\mu=+(k^{\mu}+2\delta^{\mu0}\mu)v\sqrt2\re\psi,\quad
j_2^\mu=-(k^{\mu}+2\delta^{\mu0}\mu)v\sqrt2\im\psi.
$$
We can immediately see that the isospin density rotates in the isospin plane
$(1,2)$ i.e., the plane wave is circularly polarized. In this way, a~single
Goldstone boson exploits two broken-symmetry generators, as suggested by the
general form of the commutator \eqref{charge_order_parameter}. It is notable
that the plane wave with the opposite circular polarization corresponds to the
gapped excitation in the sector $(\pi_1,\pi_2)$.

The remaining two currents are conveniently expressed in the rotated basis,
explicitly separating the unbroken and broken generator,
\begin{align*}
&\tfrac12(1+\tau_3):\quad  j^{\mu}=2(k^{\mu}+\delta^{\mu0}\mu)|\psi|^2,\\
&\tfrac12(1-\tau_3):\quad  j^{\mu}=\delta^{\mu0}\mu v^2.
\end{align*}
It is seen that the isospin wave is associated with a~uniform current of the
unbroken symmetry that is, the Goldstone boson carries the unbroken charge.
This seems to be a~generic feature of type-II Goldstone bosons.

Finally, the broken generator $\tfrac12(1-\tau_3)$ gives rise just to nonzero
charge density and, moreover, is independent of the amplitude and momentum of
the isospin wave. It is therefore to be interpreted as just a~background on
which the isospin waves propagate.

\subsection{Linear sigma model for $\mathrm{SU(3)}$ sextet}
\label{Sec:lsm_for_sextet} As a~nontrivial example of a~spontaneously broken
symmetry with nonzero charge densities the linear sigma model for an
$\mathrm{SU(3)}$ sextet scalar field will now be investigated.

The Lagrangian reads
\begin{equation}
\LA=\Tr(D_{\mu}\he\Phi
D^{\mu}\Phi)-M^2\Tr\he\Phi\Phi-a\Tr(\he\Phi\Phi)^2-b(\Tr\he\Phi\Phi)^2,
\label{sextet_Lagrangian}
\end{equation}
and is invariant under the global $\mathrm{SU(3)}\times\mathrm{U(1)}$ symmetry
that transforms the scalar field $\Phi$ as $\Phi\to U\Phi\tr U$.
A~$\mathrm{U(1)}$ chemical potential is introduced so that the covariant
derivative is $D_0\Phi=(\de_0-2\imag\mu)\Phi$.

This model provides a~phenomenological description of the color-superconducting
phase of QCD with a~color-sextet pairing of quarks of a~single flavor, which
was proposed in Ref. \pubcite1. The global $\mathrm{SU(3)}$ symmetry is what
remains of the color gauge invariance after the gluons have been `integrated
out', while the $\mathrm{U(1)}$ corresponds to the baryon number. The scalar
field $\Phi$ is an effective composite field for the quark Cooper pairs.

\begin{figure}
\begin{center}
\scalebox{0.8}{\input{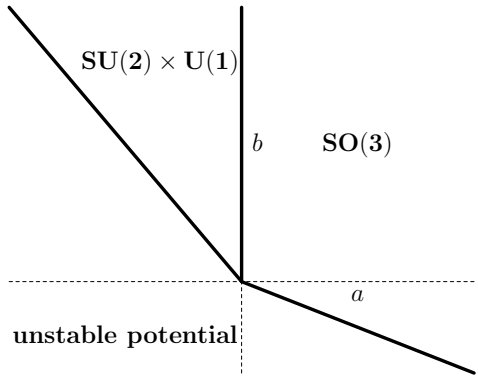}}
\end{center}
\caption{Phase diagram of the model defined by the Lagrangian
\eqref{sextet_Lagrangian}. The ordered phases are labeled by the symmetry of
the ground state. The `unstable potential' region marks a~domain of parameters
where the tree-level potential is not bounded from below.}
\label{Fig:sextet_phase_diagram}
\end{figure}

It turns out that this theory has two different ordered phases, with different
symmetry-breaking patterns and excitation spectra, see Fig.
\ref{Fig:sextet_phase_diagram}. The Bose--Einstein condensation sets at
$\mu=M/2$. All phase transitions, between the normal and an ordered phase as
well as between the ordered phases, are of second order.

In general, the excitations are grouped into multiplets of the unbroken
symmetry. This means that the more of the original
$\mathrm{SU(3)}\times\mathrm{U(1)}$ symmetry is spontaneously broken, the more
complicated the structure of the spectrum is. Both phases will now be treated
separately.

\subsubsection{The $a>0$ phase}
The static part of the Lagrangian \eqref{sextet_Lagrangian} is minimized by
a~scalar field proportional to the unit matrix i.e., $\Phi=\Delta\openone$. The
$\mathrm{SU(3)}\times\mathrm{U(1)}$ symmetry is thus spontaneously broken to
its $\mathrm{SO(3)}$ subgroup.

With this symmetry-breaking pattern in mind, the scalar field $\Phi$ is
parametrized as
$$
\Phi(x)=e^{2\imag\theta(x)}V(x)[\Delta\openone+\vp(x)]\tr V(x).
$$
Here $\theta$ is the Goldstone boson of the spontaneously broken
$\mathrm{U(1)}$ and $V=e^{\imag\pi_k\lambda_k}$, $k=1,3,4,6,8$, contains the
$5$-plet of Goldstone bosons of the coset $\mathrm{SU(3)}/\mathrm{SO(3)}$. The
real symmetric matrix $\vp$ represents six heavy `radial' modes.

Using the notation $\Pi=\pi_k\lambda_k$, the excitation spectrum is determined
by the bilinear Lagrangian,
\begin{multline*}
\LA_{\text{bilin}}=12\Delta^2(\partial_{\mu}\theta)^2+4\Delta^2\Tr(\partial_{\mu}\Pi)^2
+\Tr(\partial_{\mu}\varphi)^2-\\
-4\Delta^2\left[a\Tr\varphi^2+b(\Tr\varphi)^2\right]
-16\mu\Delta\left[\partial_0\theta\Tr\varphi+\Tr(\varphi\partial_0\Pi)\right].
\end{multline*}

\begin{figure}
\begin{center}
\scalebox{0.8}{\input{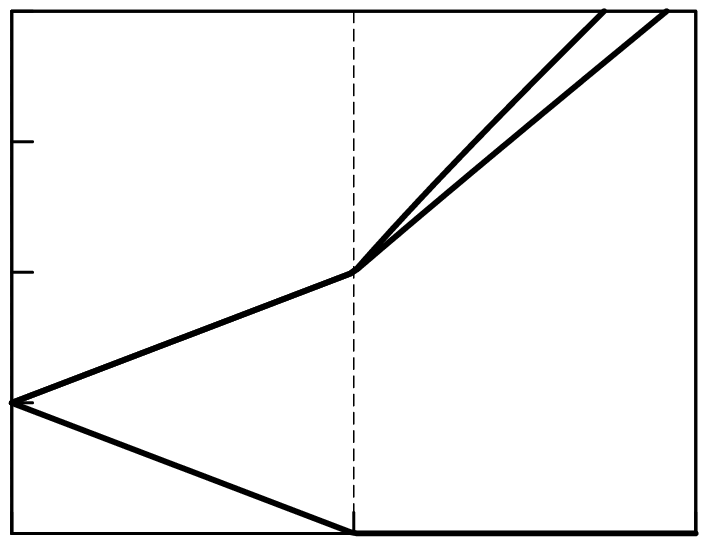}}
\end{center}
\caption{Masses of the excitations as a~function of the chemical potential in
the $\mathrm{SO(3)}$-symmetric phase. Degeneracies of the excitation branches
are indicated by the numbers. The numerical data were obtained with $a=b=1$.}
\label{Fig:sextet_spectrum1}
\end{figure}

We find that there are six Goldstone bosons, all with linear dispersion
relation. Since there are six broken generators as well, this result is in
accord with the Nielsen--Chadha counting rule. All excitations fall into
irreducible representations of the unbroken $\mathrm{SO(3)}$ group. In
particular, there is a~Goldstone singlet and a~gapped singlet in the sector
$(\theta,\Tr\vp)$. In addition, there are two $5$-plets, a~gapless and a~gapped
one, stemming from mixing of $\Pi$ with the traceless part of $\vp$, see Fig.
\ref{Fig:sextet_spectrum1}.

\subsubsection{The $a<0$ phase}
In this case the minimum of the static potential can be recast to the diagonal
form with a~single nonzero entry, $\Phi=\mathrm{diag}(0,0,\Delta)$. The
symmetry-breaking pattern is now
$\mathrm{SU(3)}\times\mathrm{U(1)}\to\mathrm{SU(2)}\times\mathrm{U(1)}$. The
scalar sextet is conveniently parametrized as
$$
\Phi(x)=e^{\imag\Pi(x)}\left(
\begin{array}{cc}
\sigma(x) & \\
& \Delta+H(x)
\end{array}\right)e^{\imag\tr\Pi(x)}.
$$
The matrix field $\Pi$ is again given by the linear combination of the broken
generators, $\Pi=\pi_k\lambda_k$, $k=4,5,6,7,8$, $\sigma$ is a~complex
symmetric $2\times2$ matrix, and $H$ is a~real scalar.

\begin{figure}
\begin{center}
\scalebox{0.8}{\input{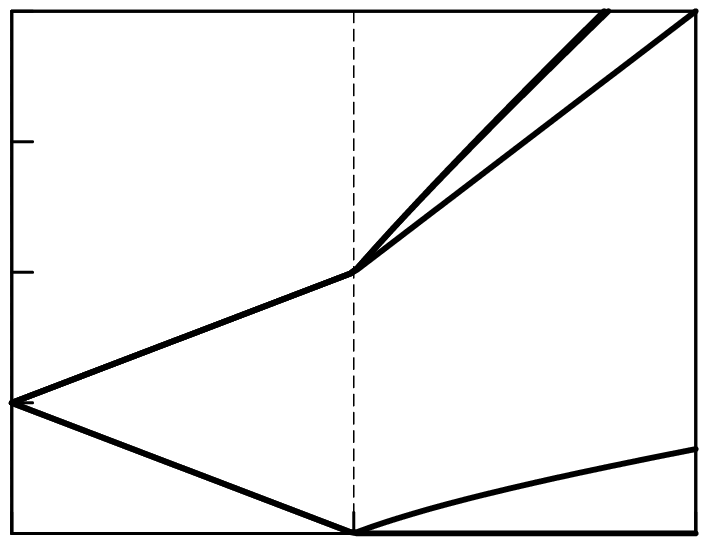}}
\end{center}
\caption{Masses of the excitations as a~function of the chemical potential in
the $\mathrm{SU(2)}\times\mathrm{U(1)}$ symmetric phase. Degeneracies of the
excitation branches are indicated by the numbers. The numerical data were
obtained with $a=-0.5$ and $b=1$.} \label{Fig:sextet_spectrum2}
\end{figure}

The bilinear part of the Lagrangian is
\begin{multline*}
\LA_{\text{bilin}}=\Tr(\partial_{\mu}\he\sigma\partial^{\mu}\sigma)+(\partial_{\mu}H)^2+
2\Delta^2(\partial_{\mu}\Pi\partial^{\mu}\Pi)_{33}+2\Delta^2(\partial_{\mu}\Pi_{33})^2-\\
-4\Delta^2(a+b)H^2+2\Delta^2a\Tr\he\sigma\sigma-16\mu\Delta
H\partial_0\Pi_{33}-
4\mu\Delta^2\im[\Pi,\partial_0\Pi]_{33}-4\mu\im\Tr\he\sigma\partial_0\sigma.
\end{multline*}
The $\mathrm{SU(2)}$ singlets $H$ and $\pi_8$ mix, giving a~Goldstone boson
with linear dispersion law and a~massive `radial' mode. The fields
$\pi_4,\pi_5,\pi_6,\pi_7$ altogether form a~complex doublet of
$\mathrm{SU(2)}$. They yield a~doublet of gapped modes and a~doublet of type-II
Goldstone bosons with a~quadratic dispersion relation. Finally, the complex
matrix $\sigma$ contains two real triplets of massive particles. For summary
see Fig. \ref{Fig:sextet_spectrum2}.

Note that there are now only three Goldstone bosons even though five generators
are spontaneously broken. This is, however, again in agreement with the
Nielsen--Chadha rule since two of the Goldstones are of the second type. Their
existence is connected with the fact that in this case, the generator
$\lambda_8$ develops nonzero ground-state density. The modified Goldstone boson
counting suggested by Eq. \eqref{charge_order_parameter} thus applies.

\subsubsection{Phase boundary}
At the boundary between the two ordered phases the model displays quite
remarkable properties. The Lagrangian \eqref{sextet_Lagrangian} is then
invariant under an extended $\mathrm{SU(6)}\times\mathrm{U(1)}$ symmetry under
which $\Phi$ transforms as a~fundamental sextet. The minima of the potential
corresponding to the two phases are now degenerate and both leave unbroken the
$\mathrm{SU(5)}\times\mathrm{U(1)}$ subgroup meaning that there are altogether
eleven broken generators.

This enhanced symmetry must, of course, be reflected in the number and type of
the Goldstone bosons \cite{Sannino:2001fd}. Indeed, by properly performing the
limit $a\to0$ it can be shown on both sides of the phase transition that there
are six Goldstone bosons. One is an $\mathrm{SU(5)}$ singlet and has a~linear
dispersion law -- this is the superfluid phonon. The other five transform as
the fundamental $\mathrm{SU(5)}$ $5$-plet and all have a~quadratic dispersion
that is, are type-II. The Nielsen--Chadha counting is thus saturated as
expected.

\subsection{General analysis}
The results achieved so far by the study of linear sigma models with particular
symmetries will now be extended to the general case. We start with the
formulation and a~short discussion of our main result: \emph{Nonzero vacuum
density of a~commutator of two broken generators implies the existence of one
type-II Goldstone boson with a~quadratic dispersion law.}

The existence of a~single Goldstone boson corresponding to two broken
generators, whose commutator has nonzero density, has been expected on the
basis of Eq. \eqref{charge_order_parameter}. Here we explicitly prove the
missing piece that is, the Goldstone boson is type-II as it must be in order to
satisfy the Nielsen--Chadha counting rule. We shall also see that the statement
formulated above holds strictly speaking only when a~convenient basis of broken
generators is chosen.

In a~sense, this result is converse to the theorem by Schaefer et al.
\cite{Schaefer:2001bq}. While they prove that zero density of commutators of
broken charges implies usual counting of the Goldstone bosons, here we show
that nonzero densities, on the contrary, lead to the existence of type-II
Goldstones and thus modified counting.

Let us consider the linear sigma model with chemical potential assigned to one
or more generators of the internal symmetry group. In general, the chemical
potential for a~conserved charge $Q$ is introduced by replacing the Hamiltonian
$H$ with $H-\mu Q$. The key observation is that, as far as \emph{exact}
symmetry is concerned, the chemical potential is always assigned to
a~$\mathrm{U(1)}$ factor of the symmetry group that is, the charge $Q$ commutes
with all generators of the exact symmetry group. The reason is that even if the
charge $Q$ is originally a~part of some larger non-Abelian symmetry group, by
adding it to the Hamiltonian we explicitly break all generators that do not
commute with it.

The Lagrangian for the general linear sigma model is defined as
\begin{equation}
\LA=D_{\mu}\he\phi D^{\mu}\phi-V(\phi).
\label{general_sigma_Lagrangian}
\end{equation}
The scalar field $\phi$ transforms under a~given representation of the global
symmetry group $\mathrm G$ and $V(\phi)$ is the most general $\mathrm
G$-invariant renormalizable potential. Finally the chemical potential enters
the Lagrangian through the covariant derivative $D_{\mu}\phi=(\de_{\mu}-\imag
A_{\mu})\phi$ \cite{Kapusta:1981aa}, $A_{\mu}$ being the constant external
gauge field which is eventually set to $A_{\mu}=(\mu Q,0,0,0)$ or the sum of
similar terms, when more chemical potentials are present.

The presence of the chemical potential destabilizes the perturbative ground
state, $\phi=0$, and eventually leads to spontaneous symmetry breaking by the
Bose--Einstein condensation. We assume that the new minimum $\phi_0$ breaks the
global symmetry group of the Lagrangian, $\mathrm G$, to its subgroup $\mathrm
H$. All generators, both broken and unbroken, are then classified by
irreducible representations of $\mathrm H$.

In the spontaneously broken phase the scalar field is parametrized as
\begin{equation}
\phi(x)=e^{\imag\Pi(x)}\left[\phi_0+H(x)\right].
\label{general_parametrization}
\end{equation}
The matrix $\Pi$ is a~linear combination of the broken generators while $H$
contains the massive (Higgs) fields. Upon expanding the Lagrangian
\eqref{general_sigma_Lagrangian} in terms of the field components, its bilinear
part becomes
\begin{multline}
\LA_{\text{bilin}}=\partial_{\mu}\he
H\partial^{\mu}H-V_{\text{bilin}}(H)-2\im\he
HA^{\mu}\partial_{\mu}H+\\
+\he\phi_0\partial_{\mu}\Pi\partial^{\mu}\Pi\phi_0-4\re\he
HA^{\mu}\partial_{\mu}\Pi\phi_0-\im\he\phi_0
A^{\mu}[\Pi,\partial_{\mu}\Pi]\phi_0.
\label{general_sigma_bilinear}
\end{multline}
Here $V_{\text{bilin}}$ is the bilinear part of the potential, which involves
only the `radial' field $H$, due to the used parametrization
\eqref{general_parametrization}.

Eq. \eqref{general_sigma_bilinear} is the main result which contains
essentially all information about the spectrum of the sigma model. To
understand better its consequences, we resort for a~moment to a~simple bilinear
Lagrangian with just two scalar fields,
\begin{equation}
\LA_{\text{bilin}}=\frac12(\partial_{\mu}\pi)^2+\frac12(\partial_{\mu}h)^2-\frac12f^2(\mu)h^2-
g(\mu)h\partial_0\pi. \label{eq:bilinear_Lagrangian}
\end{equation}
One of the fields, $h$, possibly has a~mass term and there is also
a~single-derivative mixing term, both depending explicitly on the chemical
potential. This is the generic form of the bilinear Lagrangian we met in the
two particular examples in the preceding sections.

A~simple calculation reveals that the Lagrangian \eqref{eq:bilinear_Lagrangian}
describes a~(massive) particle with dispersion relation $E^2(\vek
k)=f^2(\mu)+g^2(\mu)+\mathcal O(\vek k^2)$, and a~gapless mode with dispersion
\begin{equation}
E^2(\vek k)=\frac{f^2(\mu)}{f^2(\mu)+g^2(\mu)}\vek
k^2+\frac{g^4(\mu)}{\left[f^2(\mu)+g^2(\mu)\right]^3}\vek k^4+\mathcal O(\vek
k^6). \label{generic_GB_dispersion}
\end{equation}
If $f(\mu)=0$ that is, if both $\pi$ and $h$ are Goldstone fields mixed by the
single-derivative term, we arrive at one type-II Goldstone boson. The expansion
of its energy in powers of momentum starts at the order $\vek k^2$. On the
other hand, when $\abs{f(\mu)}>0$, the field $h$ represents a~massive mode. The
mixing of $h$ and $\pi$ then results in a~type-I Goldstone boson with linear
dispersion relation.

We can now understand the content of Eq. \eqref{general_sigma_bilinear}. There
are kinetic terms for both the radial fields $H$ and the Goldstones $\Pi$, and
the mass term for $H$, essentially given by the curvature of the static
potential at the minimum $\phi_0$. Finally, there are three mixing terms with
a~single derivative, proportional to the external field $A^{\mu}$.

The analysis of the model Lagrangian \eqref{eq:bilinear_Lagrangian} tells us
that mixing of a~radial field with a~Goldstone field gives rise to one type-I
Goldstone boson. The mixing of two Goldstone fields, on the other hand,
produces one type-II Goldstone boson. A~short glance at the last term on the
right hand side of Eq. \eqref{general_sigma_bilinear} shows that the
Goldstone--Goldstone mixing term is, as expected, proportional to the
ground-state expectation value of a~commutator of two broken generators. We
have thus established the desired result that nonzero density of a~commutator
of two broken generators gives rise to a~single type-II Goldstone boson.

In order for the conclusions just reached to be reliable, we have to show that
the results of the analysis of the simple Lagrangian
\eqref{eq:bilinear_Lagrangian} are applicable to the much more complicated case
of Eq. \eqref{general_sigma_bilinear}. A~detailed proof may be found in Ref.
\pubcite3 and will not be repeated here. Instead, we limit our discussion to
a~simplified version where, nevertheless, all the essential steps are provided.

The crucial observation regarding the charge densities is that one may always
choose a~basis of broken generators so that all generators with nonzero vacuum
expectation value mutually commute. We give a~simple proof of this statement
for the case of unitary symmetries \cite{Buballa:2005bv,Rajagopal:2005dg}. The
set of vacuum expectation values $\bra0 Q_a\ket0$ of the generators may by
regarded as a~vector $v_a$ in the space of the adjoint representation of the
Lie algebra $\lie g$ of the group $\mathrm G$. In the fundamental
representation of the unitary group, the generators $Q_a$ are realized by
Hermitian matrices, say $T_a$. Now $v_aT_a$ is also a~Hermitian matrix and as
such can be diagonalized by a~proper unitary transformation. After this
transformation $v_aT_a$ is a~linear combination of just the diagonal generators
of the symmetry group that all mutually commute i.e., span the Cartan
subalgebra of $\lie g$.

We can now take up the generators that have nonzero density in the ground state
and complement them to the Cartan subalgebra of $\lie g$. The rest of the
generators is grouped according to the standard root decomposition of Lie
algebras \cite{Georgi:1982jb}. The point is that within this basis, for any
generator there is a~unique generator such that their commutator lies in the
Cartan subalgebra. It is now proved that the broken generators participate in
the last term of Eq. \eqref{general_sigma_bilinear} in pairs and the simple
two-field analysis of Eq. \eqref{eq:bilinear_Lagrangian} is therefore
applicable.

It should, of course, also be proved that the same conclusion is true for the
mixing of the Goldstone fields with the radial ones, and of the radial ones
with themselves. Omitting the details, we just note that this follows from the
Wigner--Eckart theorem upon a~proper decomposition of the matrix fields $\Pi$
and $H$ into irreducible representations of the unbroken subgroup $\mathrm H$.


\chapter{Dynamical electroweak symmetry breaking}
\label{Chap:EWSB} The standard model of electroweak interactions has been one
of the most successful achievements of modern physics. Within a~simple and
elegant framework, it perfectly describes essentially all experimental data
collected so far. It is, however, somewhat disturbing that its only ingredient
that has not been experimentally verified yet, the Higgs boson, is crucial for
the mechanism of symmetry breaking of the
$\mathrm{SU(2)_L}\times\mathrm{U(1)_Y}$ gauge invariance and thus also for the
generation of the masses of the elementary particles. Anyway, arguments based
on the naturalness principle suggest that the standard model is just
a~low-energy limit of some more fundamental theory, and that new physics is
most likely to be found at the energies accessible already to the upcoming LHC
machine at CERN.

In this Chapter we shall take a~different point of view of the standard model.
In Section \ref{Sec:Linear_sigma_model} we explained how a~phenomenological
Lagrangian of the Ginzburg--Landau type may be used to induce spontaneous
symmetry breaking. We have, however, emphasized that such an approach is
physically unsatisfactory since it does not give an answer to the basic
question about the origin of symmetry breaking.

This happens exactly in the standard model, where the scalar sector is
introduced for sake of breaking the gauge symmetry. Attempts at replacing the
conventional Higgs mechanism with a~dynamical model of electroweak symmetry
breaking appeared soon after the construction of the standard model itself
\cite{Weinberg:1979bn,Susskind:1979ms}. The introduction to the idea of
dynamical electroweak symmetry breaking may be found in the lecture notes
\cite{Chivukula:1996uy,Lane:2002wv}, while a~more detailed review is provided
by Refs. \cite{King:1994yr,Hill:2002ap}.

The technicolor scenarios dispose with the elementary Higgs and, instead of its
vacuum expectation value, generate the order parameter for symmetry breaking by
a~fermion--antifermion condensate. This is bound together by a~new strong gauge
interaction.

Here we propose a~different idea for dynamical electroweak symmetry breaking.
We retain the elementary scalar, but with a~positive mass squared so that the
usual particle interpretation is preserved even in the absence of interactions.
Our basic assumption is the existence of a~strong Yukawa interaction between
the scalar and the massless fermions. We show that, provided the Yukawa
coupling is large enough, the fermion masses may be generated spontaneously as
a~self-consistent solution of the Schwinger--Dyson equations.

In other words, no strong gauge force is needed. The strong Yukawa interaction
breaks spontaneously the chiral symmetry, allowing for nonzero fermion masses.
Only after then, the $\mathrm{SU(2)_L}\times\mathrm{U(1)_Y}$ gauge interaction
is switched on perturbatively, resulting in the same symmetry-breaking pattern
as in the Higgs mechanism.

In order to make the proposed mechanism more transparent, we first demonstrate
it on the dynamical breaking of a global Abelian chiral symmetry, following our
paper \pubcite2. The concluding section is devoted to the discussion of the
extension to the full $\mathrm{SU(2)_L}\times\mathrm{U(1)_Y}$ gauge symmetry
\pubcite5. This model, as well as the Abelian one with the axial symmetry
gauged, are, however, still being worked on.

\section{Toy model: Global Abelian chiral symmetry}
\label{Sec:Abelian_model} We consider a~model of two Dirac fermions and
a~complex scalar defined by the Lagrangian,
\begin{multline}
\LA=\sum_{j=1,2}\left(\bar\psi_{jL}\imag\slashed{\partial}\psi_{jL}+
\bar\psi_{jR}\imag\slashed{\partial}\psi_{jR}\right)+
\partial_{\mu}\he\phi\partial^{\mu}\phi-M^2\he\phi\phi-\lambda(\he\phi\phi)^2+\\
+y_1\left(\bar\psi_{1L}\psi_{1R}\phi+\bar\psi_{1R}\psi_{1L}\he\phi\right)+
y_2\left(\bar\psi_{2R}\psi_{2L}\phi+y_2\bar\psi_{2L}\psi_{2R}\he\phi\right).
\label{Abelian_Lagrangian}
\end{multline}
The Yukawa couplings $y_1,y_2$ are, without lack of generality, assumed to be
real. Note that this Lagrangian has a~global
$\mathrm{U(1)_{V1}}\times\mathrm{U(1)_{V2}}\times\mathrm{U(1)_{A}}$ symmetry.
The vector $\mathrm{U(1)}$'s correspond to independent phase transformations of
the two Dirac spinors $\psi_1,\psi_2$. The axial $\mathrm{U(1)}$ consists of
simultaneous transformations of all the fields concerned,
$$
\psi_1\to e^{+\imag\theta\gamma_5}\psi_1,\quad \psi_2\to
e^{-\imag\theta\gamma_5}\psi_2,\quad \phi\to e^{-2\imag\theta}\phi.
$$

Note that the scalar field $\phi$ carries the axial charge. It plays a~crucial
role in the proposed mechanism of chiral (or axial) symmetry breaking. Also,
the axial charges of the fermions are opposite in order to remove the anomaly
in the axial current. It should be stressed that, as far as global symmetry is
concerned, the axial anomaly is nothing disastrous and, in fact, gives rise to
physical effects such as the $\pi^0\to\gamma\gamma$ decay in QCD. However,
having in mind the future application to electroweak interactions where the
symmetry is gauged, we choose to remove the anomaly from the very beginning.

\subsection{Ward identities: general}
The first step in the investigation of the model \eqref{Abelian_Lagrangian} is
the analysis of the symmetry. In quantum field theory, this is encoded into a
set of Ward identities for the Green's functions. Since the existence of
a~Goldstone boson is a~robust prediction of the Goldstone theorem, we show that
the Ward identities alone provide a~lot of information about the Goldstone
boson properties. We work them out without any further dynamical assumption so
that we are later able to compare dynamical symmetry breaking with the
conventional Higgs mechanism as presented in Section
\ref{Sec:Linear_sigma_model}.

The $\mathrm{U(1)_{V1}}\times\mathrm{U(1)_{V2}}\times\mathrm{U(1)_{A}}$
symmetry of the Lagrangian implies the existence of three conserved currents,
two vector and one axial, given by
\begin{equation}
\begin{split}
&j^{\mu}_{V1}=\bar\psi_1\gamma^{\mu}\psi_1,\quad
j^{\mu}_{V2}=\bar\psi_2\gamma^{\mu}\psi_2,\\
j^{\mu}_A&=\bar\psi_1\gamma^{\mu}\gamma_5\psi_1-\bar\psi_2\gamma^{\mu}\gamma_5\psi_2+
2\imag\left[(\partial^{\mu}\phi)^{\dagger}\phi-\phi^{\dagger}\partial^{\mu}\phi\right].
\end{split}
\label{Abelian_Noether_currents}
\end{equation}
Of all the correlation functions of these currents, we shall consider the
three-point ones, with a~single current and a~pair of fermions or scalars. The
vector currents do not couple to the scalar, so there are just two non-trivial
Green's functions,
$G_{V1}^{\mu}(x,y,z)=\bra0T\{j_{V1}^{\mu}(x)\psi_1(y)\bar\psi_1(z)\}\ket0$ and
$G_{V2}^{\mu}(x,y,z)=\bra0T\{j_{V2}^{\mu}(x)\psi_2(y)\bar\psi_2(z)\}\ket0$. The
corresponding proper vertex functions $\Gamma^{\mu}_{V1,2}$ satisfy the usual
Ward identities,
$$
q_{\mu}\Gamma^{\mu}_{V1,2}(p+q,p)=S_{1,2}^{-1}(p+q)-S_{1,2}^{-1}(p),
$$
$S_{1,2}$ being the full fermion propagators.

In contrast to the vector currents, the axial current $j^{\mu}_A$ contains
a~contribution from the scalar $\phi$. As will become clear later, it is
convenient to construct a~formal scalar doublet,
$$
\Phi=\left(\begin{array}{c} \phi \\ \he\phi
\end{array}\right),
$$
and use it instead of the original scalar field $\phi$. We now introduce three
Green's functions,
$G_{A\psi_1}^{\mu}(x,y,z)=\bra0T\{j_A^{\mu}(x)\psi_1(y)\bar\psi_1(z)\}\ket0$,
$G_{A\psi_2}^{\mu}(x,y,z)=\bra0T\{j_A^{\mu}(x)\psi_2(y)\bar\psi_2(z)\}\ket0$,
and
$G_{A\phi}^{\mu}(x,y,z)=\bra0T\{j_A^{\mu}(x)\Phi(y)\Phi^{\dagger}(z)\}\ket0$.
The corresponding Ward identities read
\begin{equation}
\begin{split}
q_{\mu}\Gamma^{\mu}_{A\psi_1}(p+q,p)&=S_1^{-1}(p+q)\gamma_5+\gamma_5S_1^{-1}(p),\\
q_{\mu}\Gamma^{\mu}_{A\psi_2}(p+q,p)&=-S_2^{-1}(p+q)\gamma_5-\gamma_5S_2^{-1}(p),\\
q_{\mu}\Gamma^{\mu}_{A\phi}(p+q,p)&=-2D^{-1}(p+q)\Xi+2\Xi D^{-1}(p).
\end{split}
\label{Ward_identities}
\end{equation}
Here $\imag D(x-y)=\bra0T\{\Phi(x)\he\Phi(y)\}\ket0$ is the matrix propagator
of the scalar doublet and $\Xi$ is the diagonal matrix in the scalar doublet
space, $\Xi=\mathrm{diag}(1,-1)$.

\subsubsection{Ward identities for the Higgs mechanism} The Ward identities
\eqref{Ward_identities} must hold whether the symmetry is spontaneously broken
or not. Also, they do not depend on the particular dynamical way the symmetry
is broken. As a~warmup, we shall therefore show how they fit the tree-level
analysis of the Higgs mechanism discussed in Section
\ref{Sec:Linear_sigma_model} i.e., we assume for a~moment that $M^2<0$ in Eq.
\eqref{Abelian_Lagrangian}.

Upon the expansion of the scalar field, $\phi=(v+H+\imag\pi)/\sqrt2$, the
Yukawa interaction becomes
\begin{equation}
\LA_{\text{Yukawa}}=\sum_{j=1,2}\left(m_j\bar\psi_j\psi_j+\frac{m_j}v\bar\psi_j\psi_jH\right)+
\frac\imag
v\left(m_1\bar\psi_1\gamma_5\psi_1-m_2\bar\psi_2\gamma_5\psi_2\right)\pi,
\label{Yukawa_expanded}
\end{equation}
where $m_{1,2}=vy_{1,2}/\sqrt2$ are the generated fermion masses.

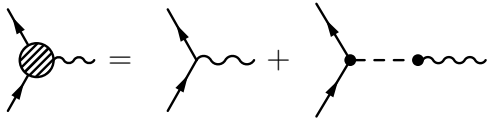
\begin{figure}
$$
\parbox{15\unitlength}{%
\begin{fmfgraph*}(15,15)
\fmfset{arrow_len}{0.16w}
\fmfsurroundn{v}{3}
\fmf{fermion}{v3,v,v2}
\fmf{photon}{v,v1}
\fmfv{d.sh=circle,d.fi=shaded,d.si=0.3w}{v}
\end{fmfgraph*}}
\,=
\parbox{15\unitlength}{%
\begin{fmfgraph*}(15,15)
\fmfset{arrow_len}{0.16w}
\fmfsurroundn{v}{3}
\fmf{fermion}{v3,v,v2}
\fmf{photon}{v,v1}
\end{fmfgraph*}}
\,+
\parbox{25\unitlength}{%
\begin{fmfgraph*}(25,15)
\fmfset{arrow_len}{0.096w}
\fmfleftn{l}{2}
\fmfright{r}
\fmf{fermion}{l1,vl,l2}
\fmf{dashes}{vl,vr}
\fmf{photon}{vr,r}
\fmfv{d.sh=circle,d.si=0.05w,d.fi=full}{vl}
\fmfv{d.sh=circle,d.si=0.05w,d.fi=full}{vr}
\end{fmfgraph*}}
$$
\caption{The axial three-point vertex function in the case of the Higgs
mechanism. The second Feynman graph on the right hand side contains a pole due
to the Goldstone boson.} \label{Fig:WI_Higgs_pion_pole}
\end{figure}

We shall exemplify the saturation of the axial Ward identity on the case of
a~fermion, say $\psi_1$. The right hand side of Eq. \eqref{Ward_identities} then
becomes
\begin{equation}
S_1^{-1}(p+q)\gamma_5+\gamma_5S_1^{-1}(p)=(\slashed p+\slashed
q+m_1)\gamma_5+\gamma_5(\slashed p+m_1)=\slashed q\gamma_5+2m_1\gamma_5.
\label{WI_RHS}
\end{equation}
The proper three-point vertex function consists, at the tree level, of two
contributions -- the bare coupling of the fermion to the axial current and
a~pion pole term, see Fig. \ref{Fig:WI_Higgs_pion_pole}.

The direct coupling of the fermion to the axial current accounts for the
$\slashed q\gamma_5$ term on the right hand side of Eq. \eqref{WI_RHS}. With
the help of Eqs. \eqref{current_pion_contribution} and \eqref{Yukawa_expanded}
we see that the pion pole contribution becomes
$$
q_{\mu}\left[-\frac{m_1}v\gamma_5\times\frac\imag{q^2}\times(2\imag
vq^{\mu})\right]=2m_1\gamma_5.
$$
Note that the last factor is $2\imag vq^{\mu}$ instead of $-\imag vq^{\mu}$, as
Eq. \eqref{current_pion_contribution} would suggest, because of a~different
normalization of the scalar contribution to the axial current
\eqref{Abelian_Noether_currents}.

We have thus verified that the axial Ward identity \eqref{Ward_identities} is
indeed satisfied. Moreover, it is now clear that, in order to compensate for
the symmetry-breaking (mass) term in Eq. \eqref{WI_RHS}, \emph{there must be
a~massless pole in the broken current correlation function due to the propagation
of the Goldstone boson}.

This observation will be crucial for the analysis of our model of dynamical
symmetry breaking. While in the Higgs mechanism (where the Goldstone boson
corresponds to an elementary field in the Lagrangian) the Ward identities serve
merely as a~check of consistency, here they will be used to predict the
properties of the composite Goldstone boson.

\subsection{Spectrum of scalars}
\label{Sec:Dynamical_symmetry_breaking} From now on we shall assume that
$M^2>0$ in the Lagrangian \eqref{Abelian_Lagrangian}, i.e., in the absence of
interactions the scalar field $\phi$ annihilates a~complex particle of mass
$M$. Our goal is to show that once a~sufficiently strong Yukawa interaction is
introduced, the axial $\mathrm{U(1)_A}$ symmetry is spontaneously broken and
fermion masses are generated.

Our strategy will be as follows: We shall \emph{assume} that fermion masses or
more precisely, chirality-changing self-energies, are somehow generated.
Plugging them into the Schwinger--Dyson equations for the Green's functions of
the theory we later show that a~nontrivial solution actually does exist. This
is a~standard philosophy in dealing with dynamical symmetry breaking -- one
simply has to make a~proper ansatz that incorporates one's expectations as to
the form of the solution.

The fermions, however, interact with the scalar $\phi$, so it is natural to
ask, and investigate prior to any calculation, what is the impact of chiral
symmetry breaking on the spectrum in the scalar sector.

\begin{figure}
$$
\parbox{30\unitlength}{%
\begin{fmfgraph*}(30,15)
\fmfset{arrow_len}{0.08w} \fmfleft{l} \fmfright{r}
\fmf{scalar,label=$\phi$,l.si=left,tension=4}{vl,l}
\fmf{scalar,label=$\phi$,l.si=right,tension=4}{vr,r} \fmf{phantom,right}{vr,vl}
\fmf{phantom,right}{vl,vr} \fmffreeze \fmfipath{p[]}
\fmfiset{p1}{vpath(__vr,__vl)} \fmfiset{p2}{vpath(__vl,__vr)}
\fmfiv{d.sh=circle,d.si=0.1w,d.fi=full}{point length(p1)/2 of p1}
\fmfiv{d.sh=circle,d.si=0.1w,d.fi=full}{point length(p2)/2 of p2}
\fmfi{fermion,label=$\psi_{1R}$,l.si=right}{subpath(0,length(p1)/2) of p1}
\fmfi{fermion,label=$\psi_{1L}$,l.si=right}{subpath(length(p1)/2,length(p1)) of
p1} \fmfi{fermion,label=$\psi_{1R}$,l.si=right}{subpath(0,length(p2)/2) of p2}
\fmfi{fermion,label=$\psi_{1L}$,l.si=right}{subpath(length(p2)/2,length(p2)) of
p2}
\end{fmfgraph*}}
\,+\,
\parbox{30\unitlength}{%
\begin{fmfgraph*}(30,15)
\fmfset{arrow_len}{0.08w} \fmfleft{l} \fmfright{r}
\fmf{scalar,label=$\phi$,l.si=left,tension=4}{vl,l}
\fmf{scalar,label=$\phi$,l.si=right,tension=4}{vr,r} \fmf{phantom,right}{vr,vl}
\fmf{phantom,right}{vl,vr} \fmffreeze \fmfipath{p[]}
\fmfiset{p1}{vpath(__vr,__vl)} \fmfiset{p2}{vpath(__vl,__vr)}
\fmfiv{d.sh=circle,d.si=0.1w,d.fi=full}{point length(p1)/2 of p1}
\fmfiv{d.sh=circle,d.si=0.1w,d.fi=full}{point length(p2)/2 of p2}
\fmfi{fermion,label=$\psi_{2L}$,l.si=right}{subpath(0,length(p1)/2) of p1}
\fmfi{fermion,label=$\psi_{2R}$,l.si=right}{subpath(length(p1)/2,length(p1)) of
p1} \fmfi{fermion,label=$\psi_{2L}$,l.si=right}{subpath(0,length(p2)/2) of p2}
\fmfi{fermion,label=$\psi_{2R}$,l.si=right}{subpath(length(p2)/2,length(p2)) of
p2}
\end{fmfgraph*}}
$$
\caption{One-loop contributions to the anomalous scalar proper self-energy. The
solid blobs denote the full chirality-changing fermion self-energies.}
\label{Fig:phi_phi_graphs}
\end{figure}
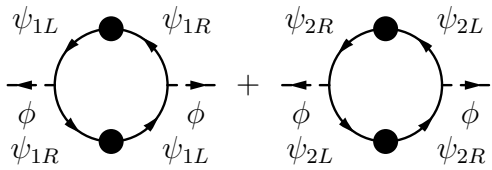

The answer lies in the fact that the scalar field carries nonzero axial charge.
Once the axial $\mathrm{U(1)_A}$ is spontaneously broken, the scalar field
carries no conserved quantum number and nothing prevents the appearance of the
`anomalous'\footnote{The word `anomalous' has nothing to do with the axial
anomaly. This terminology is taken over from condensed-matter physics, where it
is used e.g. for superconductors. There, the particle-number-violating Green's
function appears because of the Cooper pairing \cite{Fetter:1971fw}.} Green's
function $\bra0T\{\phi\phi\}\ket0$. In the language of the Feynman graphs, this
corresponds to diagrams with two external scalar legs, both pointing outwards.
Such graphs may only arise in the presence of nonzero fermion masses, see Fig.
\ref{Fig:phi_phi_graphs}. We can thus see that the breaking of the chiral
symmetry in the fermion sector (i.e., fermion masses) is tightly connected to
the breaking in the scalar sector.

The effect of the anomalous Green's function on the scalar spectrum may be
roughly understood by assuming that it is momentum-independent, and neglecting
all other radiative corrections to the scalar propagator. The scalar spectrum
is then determined by the bilinear Lagrangian
$$
\LA^{(0)}_{\text{scalar}}=\partial_{\mu}\he\phi\partial^{\mu}\phi-M^2\he\phi\phi
-\frac12\mu^{2*}\phi\phi-\frac12\mu^2\he\phi\he\phi,
$$
the parameter $\mu^2$ corresponding to the anomalous correlation function in
question. It turns out that such a~Lagrangian describes two \emph{real} scalar
particles with masses $M^2_{1,2}=M^2\pm\abs\mu^2$. The anomalous correlation
function thus amounts to the splitting of the spectrum in the scalar sector.

\subsection{Ward identities for dynamically broken symmetry}
As noted above, the vertex function of a broken current possesses a~massless
pole due to the corresponding Goldstone boson. With our assumption that the
axial symmetry is spontaneously broken, there must be such a~Goldstone boson
coupled to the axial current. Unlike the Higgs mechanism, however, now it is
a~composite particle i.e., a~bound state of the elementary fermions and
scalars. It again gives rise to a~pole in the vertex function, but now due to
quantum loops, see Fig. \ref{Fig:pole_vertex_functions}.
\begin{figure}
$$
\Gamma^{\mu}_{A\psi_1\text{,pole}}=
\parbox{30\unitlength}{%
\begin{fmfgraph*}(30,15)
\fmfset{arrow_len}{0.08w}
\fmfleftn{l}{2}
\fmfright{R}
\fmf{fermion,tension=2,label=$1$,l.si=right,l.di=0.03w}{l1,vl,l2}
\fmf{dbl_plain,tension=3}{vl,vc}
\fmfv{d.sh=circle,d.si=0.05w,d.fi=empty}{vl}
\fmfv{d.sh=circle,d.si=0.05w,d.fi=empty}{vc}
\fmfv{label=$1,,2$,l.an=180,l.di=0.06w}{r}
\fmf{fermion,right}{r,vc,r}
\fmf{photon,tension=3}{r,R}
\end{fmfgraph*}}
\,\,+
\parbox{30\unitlength}{%
\begin{fmfgraph*}(30,15)
\fmfset{arrow_len}{0.08w}
\fmfleftn{l}{2}
\fmfright{R}
\fmf{fermion,tension=2,label=$1$,l.si=right,l.di=0.03w}{l1,vl,l2}
\fmf{dbl_plain,tension=3}{vl,vc}
\fmfv{d.sh=circle,d.si=0.05w,d.fi=empty}{vl}
\fmfv{d.sh=circle,d.si=0.05w,d.fi=empty}{vc}
\fmf{dbl_dashes,right}{r,vc,r}
\fmf{photon,tension=3}{r,R}
\end{fmfgraph*}}
$$
\caption{The pole part of the proper vertex function of the axial current and
the fermion pair $\psi_1\bar\psi_1$. The double solid line represents the
Goldstone boson and the empty circles its effective vertices with the fermion
and the scalar, respectively. The double dashed line stands for the propagator
of the doublet scalar $\Phi$. Both $\psi_1$ and $\psi_2$ can circulate in the
closed fermion loop. The graphs for the other two vertex functions of the axial
current are analogous.} \label{Fig:pole_vertex_functions}
\end{figure}

As the Goldstone boson is composite, its interaction vertices cannot be
inferred directly from the Lagrangian. They can, however, be determined with
the help of the Ward identities \eqref{Ward_identities}, in terms of the
fermion and scalar propagators \cite{Hosek:1987gf,Margolis:1984cs}. Denoting
the proper vertex functions as $P_{\psi_1}$, $P_{\psi_2}$, $P_{\phi}$ (quite
analogously to the $\Gamma^{\mu}$'s, just the axial current is replaced with
the Goldstone boson), the resulting formulas read
\begin{equation}
\begin{split}
P_{\psi_1}(p+q,p)&=\frac1N\left[S_1^{-1}(p+q)\gamma_5+\gamma_5S_1^{-1}(p)-\slashed
q\gamma_5\right],\\
P_{\psi_2}(p+q,p)&=-\frac1N\left[S_2^{-1}(p+q)\gamma_5+\gamma_5S_2^{-1}(p)-\slashed
q\gamma_5\right],\\
P_{\phi}(p+q,p)&=-\frac2N\left[D^{-1}(p+q)\Xi-\Xi
D^{-1}(p)-q\cdot(2p+q)\Xi\right],
\end{split}
\label{effective_vertices_general}
\end{equation}
the normalization factor $N$ will be specified later. Note that these effective
vertices are unambiguous only up to order $\mathcal O(q)$ since only the pole
parts of the axial current vertex functions were kept in the Ward identities
\cite{Jackiw:1973tr}.

To determine the Goldstone interactions more concretely, the knowledge of the
full fermion and scalar propagators is necessary. It is, however, obvious that
the most important are their symmetry-breaking parts. In order to be able to
write down analytic expressions for the vertices
\eqref{effective_vertices_general}, we make the following simplifications.

We neglect the symmetry-preserving renormalization of the fermion and scalar
propagators and assume that the sheer effect of quantum corrections is to
generate the symmetry breaking so that the propagators acquire the form
\begin{equation}
S_{1,2}^{-1}(p)=\slashed p-\Sigma_{1,2}(p),\quad
D^{-1}(p)=\left(
\begin{array}{cc}
p^2-M^2 & -\Pi(p) \\
-\Pi^*(p) & p^2-M^2
\end{array}\right).
\label{ansatz_propagators}
\end{equation}
Here $\Sigma_{1,2}(p)$ are the chirality-changing proper self-energies of the
fermions while $\Pi(p)$ is the anomalous proper self-energy of the scalar field
$\phi$.

The effective vertices \eqref{effective_vertices_general} now become
\begin{equation}
\begin{split}
P_{\psi_1}(p+q,p)&=-\frac1N\left[\Sigma_1(p+q)+\Sigma_1(p)\right]\gamma_5,\quad
P_{\psi_2}(p+q,p)=\frac1N\left[\Sigma_2(p+q)+\Sigma_2(p)\right]\gamma_5,\\
P_{\phi}(p+q,p)&=-\frac2N\left(
\begin{array}{cc}
0 & \Pi(p+q)+\Pi(p) \\
-\Pi^*(p+q)-\Pi^*(p) & 0
\end{array}\right).
\label{effective_vertices}
\end{split}
\end{equation}
The normalization factor $N$ is given by
$N=\sqrt{J_{\psi_1}(0)+J_{\psi_2}(0)+J_{\phi}(0)}$, the loop integrals
$J_{\psi_1}(q^2)$, $J_{\psi_2}(q^2)$ and $J_{\phi}(q^2)$ being defined as
\begin{align*}
-\imag q^{\mu}J_{\psi_{1,2}}(q^2)&=8\int\frac{\dfour
k}{(2\pi)^4}\frac{(k-q)^{\mu}\Sigma_{1,2,k}}{k^2-\Sigma^2_{1,2,k}}
\frac{\Sigma_{1,2,k}+\Sigma_{1,2,k-q}}{(k-q)^2-\Sigma^2_{1,2,k-q}},\\
-\imag q^{\mu}J_{\phi}(q^2)&=8\int\frac{\dfour
k}{(2\pi)^4}\frac{(2k-q)^{\mu}(k^2-M^2)}{(k^2-M^2)^2-|\Pi_k|^2}
\frac{\re\left[\Pi^*_{k-q}\bigl(\Pi_{k}+\Pi_{k-q}\bigr)\right]}{\bigl[(k-q)^2-M^2\bigr]^2-|\Pi_{k-q}|^2}.
\end{align*}

\subsection{Spectrum of fermions}
\label{Sec:SD_equations} So far we have simply assumed that axial symmetry is
spontaneously broken, giving rise to nonzero proper self-energies
$\Sigma_{1,2}(p)$ and $\Pi(p)$. Now we have to close the chain of arguments by
demonstrating that this is indeed the case.

\begin{figure}
\begin{align*}
\parbox{20\unitlength}{%
\begin{fmfgraph*}(20,15)
\fmfset{arrow_len}{0.12w} \fmfleft{l} \fmfright{r} \fmf{fermion}{r,v,l}
\fmfv{d.sh=circle,d.si=0.3w,d.fi=shaded}{v}
\end{fmfgraph*}}
\,&=\,
\parbox{30\unitlength}{%
\begin{fmfgraph*}(30,20)
\fmfset{arrow_len}{0.08w} \fmfleft{l} \fmfright{r}
\fmf{fermion,tension=3}{vl,l} \fmf{fermion,tension=3}{r,vr}
\fmf{phantom,right}{vr,vl} \fmf{phantom,right}{vl,vr} \fmffreeze \fmfipath{p[]}
\fmfiset{p1}{vpath(__vr,__vl)} \fmfiset{p2}{vpath(__vl,__vr)}
\fmfi{dbl_dashes}{subpath(0,length(p1)/2) of p1}
\fmfi{dbl_dashes}{subpath(length(p1),length(p1)/2) of p1}
\fmfi{fermion}{subpath(length(p2)/2,0) of p2}
\fmfi{fermion}{subpath(length(p2),length(p2)/2) of p2}
\fmfiv{d.sh=circle,d.si=0.1w,d.fi=full}{point length(p1)/2 of p1}
\fmfiv{d.sh=circle,d.si=0.1w,d.fi=full}{point length(p2)/2 of p2}
\end{fmfgraph*}}\\
\parbox{20\unitlength}{%
\begin{fmfgraph*}(20,15)
\fmfset{arrow_len}{0.12w} \fmfleft{l} \fmfright{r} \fmf{dbl_dashes}{r,v,l}
\fmfv{d.sh=circle,d.si=0.3w,d.fi=shaded}{v}
\end{fmfgraph*}}
\,&=\,
\parbox{30\unitlength}{%
\begin{fmfgraph*}(30,15)
\fmfset{arrow_len}{0.08w} \fmfleft{l} \fmfright{r}
\fmf{dbl_dashes,tension=3}{vl,l} \fmf{dbl_dashes,tension=3}{vr,r}
\fmf{phantom,right}{vr,vl} \fmf{phantom,right}{vl,vr}
\fmfv{label=$1,,2$,l.an=180,l.di=0.13w}{vr} \fmffreeze \fmfipath{p[]}
\fmfiset{p1}{vpath(__vr,__vl)} \fmfiset{p2}{vpath(__vl,__vr)}
\fmfiv{d.sh=circle,d.si=0.1w,d.fi=full}{point length(p1)/2 of p1}
\fmfiv{d.sh=circle,d.si=0.1w,d.fi=full}{point length(p2)/2 of p2}
\fmfi{fermion}{subpath(0,length(p1)/2) of p1}
\fmfi{fermion}{subpath(length(p1)/2,length(p1)) of p1}
\fmfi{fermion}{subpath(0,length(p2)/2) of p2}
\fmfi{fermion}{subpath(length(p2)/2,length(p2)) of p2}
\end{fmfgraph*}}
\,+\,\,
\parbox{15\unitlength}{%
\begin{fmfgraph*}(15,15)
\fmfleft{l} \fmfright{r} \fmf{dbl_dashes,tension=0.4}{l,v,r}
\fmf{phantom,tension=0.4}{v,v} \fmffreeze \fmfipath{p}
\fmfiset{p}{vpath(__v,__v)} \fmfi{dbl_dashes}{subpath(0,length(p)/2) of p}
\fmfi{dbl_dashes}{subpath(length(p)/2,length(p)) of p}
\fmfiv{d.sh=circle,d.si=0.2w,d.fi=full}{point length(p)/2 of p}
\end{fmfgraph*}}
\end{align*}
\caption{The one-loop Schwinger--Dyson equations for the fermion and scalar
propagators. The first line applies equally to $\psi_1$ and $\psi_2$. The
proper self-energies are denoted by the dashed blobs, while the full
propagators are represented by the solid blobs.} \label{Fig:SD_equations}
\end{figure}
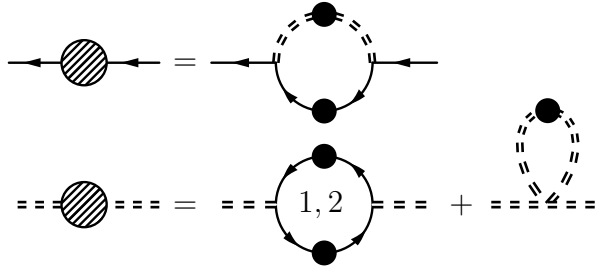

To that end, we consider the Schwinger--Dyson equations for the Green's
functions of our model. Having in mind that we are looking for spontaneous
symmetry breaking in the propagators, we neglect for simplicity all vertex
corrections. The propagators are then found by a self-consistent solution of
the one-loop equations that are depicted in Fig. \ref{Fig:SD_equations}.

With the ansatz \eqref{ansatz_propagators}, we arrive at the set of three
coupled integral equations,
\begin{equation}
\begin{split}
\Sigma_{1,p}&=\imag y_1^2\int\frac{\dfour
k}{(2\pi)^4}\frac{\Sigma_{1,k}}{k^2-\Sigma_{1,k}^2}
\frac{\Pi_{k-p}}{\left[(k-p)^2-M^2\right]^2-|\Pi_{k-p}|^2},\\
\Sigma_{2,p}&=\imag y_2^2\int\frac{\dfour
k}{(2\pi)^4}\frac{\Sigma_{2,k}}{k^2-\Sigma_{2,k}^2}
\frac{\Pi^*_{k-p}}{\left[(k-p)^2-M^2\right]^2-|\Pi_{k-p}|^2},\\
\Pi_p&=-\sum_{j=1,2}2\imag y_j^2\int\frac{\dfour
k}{(2\pi)^4}\frac{\Sigma_{j,k}}{k^2-\Sigma_{j,k}^2}
\frac{\Sigma_{j,k-p}}{(k-p)^2-\Sigma_{j,k-p}^2}+\imag\lambda\int\frac{\dfour
k}{(2\pi)^4} \frac{\Pi_k}{(k^2-M^2)^2-|\Pi_k|^2}.
\end{split}
\label{SD_model_equations}
\end{equation}

For sake of numerical solution of the Schwinger--Dyson equations
\eqref{SD_model_equations}, further simplifying assumptions are made. First,
since the symmetry-preserving quantum corrections have been neglected, we also
abandon the $\lambda$ interaction in the last of Eqs.
\eqref{SD_model_equations}. The reason is that it merely provides a~counterterm
in the one-loop effective Lagrangian, whereas the spontaneous breaking itself
is induced by the Yukawa interaction.

Second, the Yukawa couplings $y_1,y_2$ are set equal so that the set of
equations \eqref{SD_model_equations} reduces to two equations for
$\Sigma=\Sigma_1=\Sigma_2$ and $\Pi$. This conclusion is justified as long as
the scalar self-energy $\Pi$ is real, since the discrete symmetry of the
Lagrangian, $\psi_1\leftrightarrow\psi_2$ and $\phi\leftrightarrow\he\phi$, is
then not spontaneously broken.

\begin{figure}
\begin{center}
\scalebox{0.8}{\input{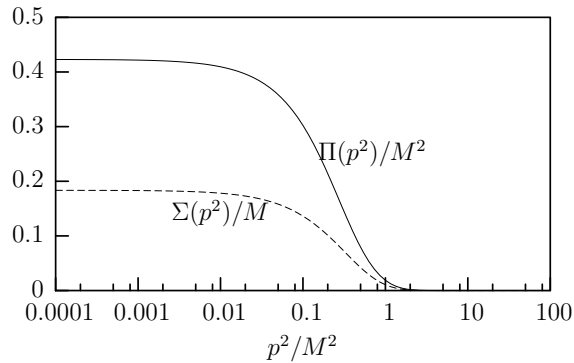}}
\end{center}
\caption{Numerical results for the fermion and scalar proper self-energies
$\Sigma$ and $\Pi$, respectively. The author is grateful to Petr Bene\v{s} for
doing the numerical computation and providing this figure.}
\label{Fig:Sigma_Pi}
\end{figure}

The numerical results of the calculations in Euclidean space are displayed in
Fig. \ref{Fig:Sigma_Pi}. It is notable that a~nontrivial solution seems to
exists only when the Yukawa interaction is strong enough. A preliminary
analysis shows that the critical value for spontaneous breaking of the chiral
symmetry is $y_{\text{crit}}\approx30$.\footnote{Very recently, we have
discovered an error in the original numerical code. Our new computations, to be
published, suggest that the critical value of the Yukawa coupling might be
significantly larger, about 80. The qualitative behavior of the self-energies,
however, does not change.}

\section{Extension to $\mathrm{SU(2)_L}\times\mathrm{U(1)_Y}$ gauge
symmetry} Having demonstrated how fermion masses may be generated dynamically
by a~strong Yukawa interaction, we now turn our attention to the case of utmost
physical importance, the spontaneous breaking of the electroweak
$\mathrm{SU(2)_L}\times\mathrm{U(1)_Y}$ gauge symmetry. This case has not been
investigated in full detail yet, including the numerical solution of the
Schwinger--Dyson equations. Therefore, we just sketch the main idea as done in
our paper \pubcite5. Further work on this model is in progress.

The basic strategy is the same as in Section \ref{Sec:Abelian_model}. The only
difference is that now all formulas are more complicated because of the isospin
and flavor structure of the standard model. The particle content is identical
to that of the standard model with two exceptions. First, in the fermion
sector, we introduce $N_f$ neutrino right-handed isospin singlets $\nu_R$ with
zero weak hypercharge in order to account for the nonzero neutrino masses.

Second, in the scalar sector, we introduce two complex doublets,
$S=(S^{(+)},S^{(0)})$ and $N=(N^{(0)},N^{(-)})$, with weak hypercharges
$Y_S=+1$ and $Y_N=-1$ and different ordinary masses $M_S$ and $M_N$,
respectively. It will become clear later that they serve to generate the masses
of the lower and upper components of the fermion isospin doublets.

The Lagrangian of our model differs from that of the standard model by the
presence of two scalar quartic self-couplings, $\lambda_S$ and $\lambda_N$, and
by the Yukawa interaction
\begin{equation}
\LA_{\text{Yukawa}}=\bar\ell_Ly_ee_RS+\bar\ell_Ly_{\nu}\nu_RN+\bar
q_Ly_dd_RS+\bar q_Ly_uu_RN+\text{H.c.}, \label{Yukawa_int}
\end{equation}
where the Yukawa couplings $y_e,y_{\nu},y_d,y_u$ are to be treated as
$N_f\times N_f$ complex matrices in the flavor space.

\subsection{Particle spectrum}
\begin{figure}
$$
{\def\figlab{e}
\parbox{30\unitlength}{%
\begin{fmfgraph*}(30,15)
\fmfkeep{loop}
\fmfset{arrow_len}{0.1w}
\fmfleft{l}
\fmfright{r}
\fmf{scalar,label=$S^{(0)}$,l.si=left,tension=4}{vl,l}
\fmf{scalar,label=$S^{(0)}$,l.si=right,tension=4}{vr,r}
\fmf{phantom,right}{vr,vl}
\fmf{phantom,right}{vl,vr}
\fmffreeze
\fmfipath{p[]}
\fmfiset{p1}{vpath(__vr,__vl)}
\fmfiset{p2}{vpath(__vl,__vr)}
\fmfiv{d.sh=circle,d.si=0.1w,d.fi=full}{point length(p1)/2 of p1}
\fmfiv{d.sh=circle,d.si=0.1w,d.fi=full}{point length(p2)/2 of p2}
\fmfi{fermion,label=$\figlab_R$,l.si=right}{subpath(0,length(p1)/2) of p1}
\fmfi{fermion,label=$\figlab_L$,l.si=right}{subpath(length(p1)/2,length(p1)) of p1}
\fmfi{fermion,label=$\figlab_R$,l.si=right}{subpath(0,length(p2)/2) of p2}
\fmfi{fermion,label=$\figlab_L$,l.si=right}{subpath(length(p2)/2,length(p2)) of p2}
\end{fmfgraph*}}
\,+\,
\def\figlab{d}
\parbox{30\unitlength}{\fmfreuse{loop}}}
$$
\caption{One-loop contributions to the anomalous proper self-energy of the
neutral scalar $S^{(0)}$. The solid blobs denote the chirality-changing parts
of the full fermion propagators. The same graphs apply to $N^{(0)}$ upon
replacing $e,d$ with $\nu,u$.} \label{Fig:S_self_energy}
\end{figure}
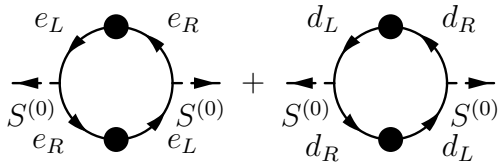

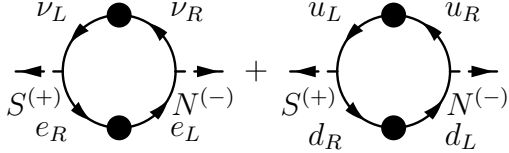
\begin{figure}
$$
{\def\figlabu{\nu}
\def\figlabd{e}
\parbox{30\unitlength}{%
\begin{fmfgraph*}(30,15)
\fmfkeep{loop_charged}
\fmfset{arrow_len}{0.1w}
\fmfleft{l}
\fmfright{r}
\fmf{scalar,label=$S^{(+)}$,l.si=left,tension=4}{vl,l}
\fmf{scalar,label=$N^{(-)}$,l.si=right,tension=4}{vr,r}
\fmf{phantom,right}{vr,vl}
\fmf{phantom,right}{vl,vr}
\fmffreeze
\fmfipath{p[]}
\fmfiset{p1}{vpath(__vr,__vl)}
\fmfiset{p2}{vpath(__vl,__vr)}
\fmfiv{d.sh=circle,d.si=0.1w,d.fi=full}{point length(p1)/2 of p1}
\fmfiv{d.sh=circle,d.si=0.1w,d.fi=full}{point length(p2)/2 of p2}
\fmfi{fermion,label=$\figlabu_R$,l.si=right}{subpath(0,length(p1)/2) of p1}
\fmfi{fermion,label=$\figlabu_L$,l.si=right}{subpath(length(p1)/2,length(p1)) of p1}
\fmfi{fermion,label=$\figlabd_R$,l.si=right}{subpath(0,length(p2)/2) of p2}
\fmfi{fermion,label=$\figlabd_L$,l.si=right}{subpath(length(p2)/2,length(p2)) of p2}
\end{fmfgraph*}}
\,+\,
\def\figlabu{u}
\def\figlabd{d}
\parbox{30\unitlength}{\fmfreuse{loop_charged}}}
$$
\caption{One-loop mixing of the charged scalars induced by the dynamically
generated fermion masses.} \label{Fig:charged_scalar_mixing}
\end{figure}

As in the simple Abelian model \eqref{Abelian_Lagrangian}, the assumed fermion
mass terms give rise to `anomalous' self-energies in the scalar sector, mixing
different modes. At one-loop, the neutral components $S^{(0)}$ and $N^{(0)}$
develop nonzero two-point correlation functions breaking the particle number,
see Fig. \ref{Fig:S_self_energy}. As a~result, there are four real particles,
two with masses split around $M_S$, and the other two around $M_N$. It should
also be noted that at higher orders, all these four modes mix with one another
since there is no conserved quantum number that would prevent them from doing
so.
The charged components of the scalar doublets also mix, as shown in Fig.
\ref{Fig:charged_scalar_mixing}. Due to the conservation of electric charge,
there are now two charged scalars, being the orthogonal mixtures of $S^{(+)}$
and $N^{(-)\dagger}$.

Leaving aside the details of the calculations that may be found in the paper
\pubcite5, we just note that the fermion and scalar self-energies are
determined as a~solution to the truncated Schwinger--Dyson equations, very much
analogous to Eqs. \eqref{SD_model_equations}.

The essential difference is that now the
$\mathrm{SU(2)_L}\times\mathrm{U(1)_Y}$ chiral symmetry is gauged i.e., the
chiral currents are coupled to dynamical vector gauge fields. As a~consequence,
the three Goldstone bosons of the coset
$[\mathrm{SU(2)_L}\times\mathrm{U(1)_Y}]/\mathrm{U(1)_Q}$ become the
longitudinal components of the three massive vector bosons $W^{\pm},Z$. The
Ward identities enable us to calculate the couplings of the Goldstone modes to
the gauge bosons. Due to the propagation of the intermediate Goldstone boson,
the self-energy of the gauge field acquires a~massless pole. Upon neglecting
the finite contributions to the polarization tensor, the gauge boson mass
squared is equal to the residue at this pole \cite{Jackiw:1973tr}. The new
feature of the proposed model is that the couplings of the Goldstones to the
gauge bosons, and hence also the gauge boson masses, are expressed through
one-loop graphs containing the symmetry-breaking self-energies of the fermions
and scalars. The gauge boson masses are therefore tied to the masses of the
other particles by certain sum rules \cite{Hosek:1987gf}.\footnote{In fact, in
the paper \pubcite5 we omitted the scalar contribution to the gauge boson
masses. Now that we have gained some experience by the study of the Abelian
model of Section \ref{Sec:Abelian_model}, the application of the idea to the
electroweak symmetry breaking is being revised.}

Finally, let us note that we have so far not dealt with the Majorana masses of
the neutrinos. It turns out that once a~hard Majorana mass term is introduced
for the right-handed neutrinos, the left-handed neutrino Majorana masses are
generated as a~one-loop effect. Together, they also produce a~new contribution
to the anomalous self-energy of $N^{(0)}$. In conclusion, it is perhaps more
appropriate to treat all the masses, both Dirac and Majorana, on the same
footing that is, self-consistently. The spectrum then contains $2N_f$ massive
Majorana neutrinos, presumably with a~seesaw-like hierarchy of masses.

\subsection{Phenomenological constraints}
Several constraints apart from reproducing the fermion and gauge boson mass
spectrum must be met before our model may be accepted as an alternative to the
standard model of electroweak interactions. Since we have not yet reached the
stage of solving the Schwinger--Dyson equations numerically, we shall discuss
these constraints only qualitatively.

First, since the Goldstone bosons of the spontaneously broken symmetry are
bound states of the elementary fermions and scalars, all the elementary scalars
remain in the spectrum of physical states, unlike in the standard model.
Consequently, the neutral ones mediate flavor-changing processes, thus
contributing to the flavor-changing neutral currents. Since these are highly
suppressed in the standard model, the scalar masses $M_S,M_N$ must be large
enough in order to avoid experimental bounds.

Second, it is well known that pair production of longitudinally polarized
massive vector bosons violates tree unitarity at high energies, rendering the
theory nonrenormalizable \cite{Chivukula:1996uy}. In order for the growth of
the scattering amplitude to be cut off at high energies, it is necessary that
there be new particles at the energy scale of order $1\,\mathrm{TeV}$.


\chapter{Quantum chromodynamics at nonzero density}
\label{Chap:QCD} The physics of hot and/or dense matter is described by the
phase diagram of QCD. While the region of low net baryon density and high
temperature is being explored experimentally in heavy ion collisions, the cold
and very dense nuclear matter seems to exist only in the neutron stars.

It has been known for a~long time that at sufficiently high density quarks are
no longer confined\footnote{In fact, the term \emph{quark confinement} loses
its sense once the mean distance between quarks is much smaller than the
confinement scale. The quarks then do not feel the long-distance strong
attraction and provide the appropriate degrees of freedom to describe the
highly squeezed matter.} and may undergo the Cooper pairing very much analogous
to that in ordinary superconductors \cite{Bailin:1984bm}. However, only in the
past decade has the phenomenon of color superconductivity attracted
considerable attention due to the discovery that it may appear already at
densities attainable in the neutron stars \cite{Rapp:1998zu,Alford:1998zt}.

Since then, the subject has been investigated to great detail and several
qualitatively different phases have been found. Extensive reviews are given in
Refs. \cite{Rajagopal:2000wf,Buballa:2003qv,Alford:2001dt,Rischke:2003mt}. An
introduction to the physics of cold dense quark matter may be found in the
lecture notes \cite{Shovkovy:2004me,Schaefer:2005ff}.

Despite the amount of energy devoted to the study of the QCD phase diagram,
there is still a~controversy regarding the structure of the ground state at
moderate baryon density. It seems that we are only confident that at very high
densities the quark matter resides in the \emph{Color-Flavor-Locked} phase
\cite{Alford:2002rz}. This is supported by the weak-coupling calculations from
first principles, which are applicable due to the asymptotic freedom of QCD.

On the other hand, the knowledge of the moderate-density region of the phase
diagram is rather weak. Usually, either the weak-coupling results are directly
extrapolated just by running the QCD coupling, or the structure of the
interaction is taken over from the high-density regime and used as an input to
the phenomenological models such as that of Nambu and Jona-Lasinio.

This chapter consists of two main parts. In the first one, we introduce an
alternative mechanism for generating the effective four-quark interaction and
show that it leads to an unconventional pairing in the color-sextet channel.
This is based on our paper \pubcite1.

The second part, based on the recent paper \pubcite4, deals with a~different
approach to the QCD phase diagram. Inasmuch as we cannot attack the problem of
the QCD phase diagram at moderate density directly and current lattice
techniques fail in that region as well, it is plausible to study theories
similar to QCD which are amenable to both analytical and lattice calculations.
We describe a~simple case of such a~theory -- the two-color QCD with two quark
flavors -- and provide a~new setting for its low-energy description in terms of
the chiral perturbation theory.

\section{Single-flavor color superconductor with color-sextet pairing}
It is most common to describe the quark matter at moderate baryon density
within the Nambu--Jona-Lasinio model \cite{Buballa:2003qv}. In this approach,
the crucial point is the choice of the model interaction. The color and flavor
structure of the interaction are usually taken over from the weak-coupling
regime -- either perturbative (the one-gluon exchange) or nonperturbative (the
instanton-mediated interaction). Both these interactions share the common
feature that they are attractive in the color-antisymmetric channel and
repulsive in the color-sym\-metric one. It should, however, be stressed that
the arguments based on the weakly coupled QCD merely provide an
\emph{evidence}. There is \emph{no proof} that the strongly coupled QCD at
moderate density inevitably leads to the same behavior. It is therefore worth
exploring the alternatives.

In this section we shall investigate the behavior of dense quark matter under
the assumption that the quarks pair in the color-symmetric (sextet) channel. We
shall for simplicity consider a~homogeneous phase of a~single-flavor quark
matter. The physical reasoning behind this assumption is the following. The
color, flavor and spin structures of the Cooper pair are connected by the
requirement that the Pauli exclusion principle be satisfied. This means that,
as long as the orbital momentum is zero, the total spin of the color-sextet
Cooper pair of quarks of a~single flavor must be zero. On the contrary, in the
color-antitriplet channel the Pauli principle requires total spin one.

The point is that the spin and orbital momentum effects dramatically reduce the
energy gap i.e., the binding energy of the Cooper pair. Indeed, while -- in the
color-antitriplet channel -- the gap of the two-flavor spin-zero superconductor
at moderate density is roughly estimated as tens $\mathrm{MeV}$, the gap of the
one-flavor spin-one superconductor is only tens or a~hundred $\mathrm{keV}$
\cite{Buballa:2002wy}. In the latter case, the color-sextet pairing might
prevail even if the pairing interaction is quite weak.

It is well known that while at very high density the CFL phase is the stable
ground state of the three-flavor quark matter, at moderate density the CFL
pairing is disfavored by the strange quark mass and the resulting mismatch of
the Fermi momenta. The $2+1$ pairing scheme is more likely. The up and down
flavors are bound by the strong attractive interaction in the color-antitriplet
channel. The strange quarks then pair with themselves and we suggest here that
the pairing be in the \emph{color-sextet spin-zero} channel rather than the
color-antitriplet spin-one channel favored by the one-gluon exchange
interaction.

We first \emph{assume} the particular form of the pairing and explore its
impact on the symmetry of the theory. It is only later that we provide
a~physical motivation for the attraction in the color-symmetric channel and work
out the description within the Nambu--Jona-Lasinio model.

\subsection{Kinematics of color-sextet condensation}
Suppose that the superconducting phase is described by the order parameter
$\Phi$ which transforms in the $\mathbf 6$ representation of the color
$\mathrm{SU(3)}$ group. It is best represented by a~complex symmetric $3\times
3$ matrix upon which the $\mathrm{SU(3)\times U(1)}$
transformations\footnote{Recall from Section \ref{Sec:lsm_for_sextet} that the
$\mathrm{U(1)}$ here represents the baryon number.} act as $\Phi\to U\Phi\tr
U$. The assumption that the ordered phase be homogeneous translates to the
requirement that $\Phi$ be a~spacetime-independent constant. Note that in the
Nambu--Jona-Lasinio model $\Phi_{ij}$ will correspond to the vacuum expectation
value of the bilinear operator $\psi_{\alpha
i}(C\gamma_5)_{\alpha\beta}\psi_{\beta j}$, but for now this interpretation is
not needed.

The crucial observation is that any complex symmetric matrix $\Phi$ may be
brought by a~suitable $\mathrm{SU(3)\times U(1)}$ transformation to a~special
form $\Delta$ which is diagonal, real and positive \cite{Schur:1945ab}. We
shall denote its diagonal entries as $\Delta_1,\Delta_2,\Delta_3$. These cannot
be changed by a~unitary transformation since they are the eigenvalues of the
positive Hermitian matrix $(\he\Phi\Phi)^{1/2}$, and thus represent \emph{three
independent order parameters} of the phase.

The presence of three order parameters makes the phase structure of the
color-sextet superconductor quite rich. Depending on the relative values of the
order parameters, several symmetry-breaking patterns may be distinguished:
\begin{enumerate}
\item \emph{All $\Delta$'s are different and nonzero.} This is the most general as
well as intriguing possibility. The continuous $\mathrm{SU(3)\times U(1)}$
symmetry is completely broken, only a~discrete $(Z_2)^3$ is left.
\item \emph{Two $\Delta$'s are equal and nonzero.} In this case, there is
a~residual $\mathrm{O(2)}$ symmetry in the corresponding $2\times2$ block of
$\Phi$.
\item \emph{$\Delta_1=\Delta_2=\Delta_3\neq0$.} Quite similar to the previous
case, but now the enhanced symmetry of the ground state is $\mathrm{O(3)}$.
\item \emph{Some of the $\Delta$'s are zero.} According to the number of
vanishing order parameters, there is a~residual $\mathrm{U(1)}$ or
$\mathrm{U(2)}$ invariance, simply meaning that the corresponding colors do not
participate in the pairing.
\end{enumerate}

It will turn out in the following that the possibility of most interest is the
$\mathrm{O(3)}$-symmetric phase. Since this results in the same number of
broken color generators as the breaking $\mathrm{SU(3)\to SU(2)}$ by the
standard color antitriplet, it is worthwhile to comment on the difference
between these two symmetry-breaking patterns.

The structure of the spectrum is always determined by the unbroken subgroup.
Now the breaking $\mathrm{SU(3)\to SO(3)}$ is isotropic so that all five broken
generators fall into a~single ($5$-plet) representation of $\mathrm{SO(3)}$. On
the other hand, in the $\mathrm{SU(3)\to SU(2)}$ case four of the broken
generators form a~complex $\mathrm{SU(2)}$ doublet while the remaining one is
a~singlet.

\subsection{Ginzburg--Landau description}
To determine which of the possible symmetry-breaking patterns are actually
realized, one has to employ a~particular model to calculate the order parameter
$\Phi$. Ignoring for the moment the fluctuations of the order parameter(s), we
have to write down the most general $\mathrm{SU(3)\times U(1)}$ invariant
potential, whose minimum determines the ground state. Such a~potential can
always be written in terms of a~certain set of algebraically independent
invariants. In our case there are three of them, namely $\Tr\he\Phi\Phi$,
$\det\he\Phi\Phi$ and $\Tr(\he\Phi\Phi)^2$.

Restricting to quartic polynomials of the Ginzburg--Landau type, the most
general potential reads
$$
V(\Phi)=-a\Tr\he\Phi\Phi+b\Tr(\he\Phi\Phi)^2+c(\Tr\he\Phi\Phi)^2.
$$
Such a~potential was already investigated in Section \ref{Sec:lsm_for_sextet}.
It was shown that the nature of the global minimum depends on the sign of the
parameter $b$. When $b>0$, the order parameter $\Delta$ is proportional to the
unit matrix so that the ground state has the $\mathrm{SO(3)}$ symmetry. When
$b<0$, the minimizing configuration is such that $\Delta$ has a~single nonzero
diagonal entry, corresponding to the symmetry breaking pattern
$\mathrm{SU(3)\times U(1)\to SU(2)\times U(1)}$.

We stress the fact that the parameters $a,b,c$ are unknown at this stage so
that we cannot decide which of the ordered phases is actually realized. It is,
however, possible to derive the Ginzburg--Landau functional from the underlying
microscopic model, either QCD or Nambu--Jona-Lasinio \cite{Giannakis:2001wz}.

To account for the fluctuations of the order parameter $\Phi$, the
Ginzburg--Landau functional has to be enriched with derivative terms. The
lowest-order Lagrangian reads
\begin{equation}
\LA
=\alpha_e\Tr\he{\de_0\Phi}\de^0\Phi+\alpha_m\Tr\he{\de_i\Phi}\de^i\Phi-V(\Phi).
\label{GL_effective_Lagr}
\end{equation}
The coefficients $\alpha_e$ and $\alpha_m$ are in general different since
Lorentz invariance is broken by medium effects. Note that the ``kinetic term''
of $\Phi$ is not canonically normalized -- this is because $\Phi$ represents a
composite object, the Cooper pair of quarks
\cite{Pisarski:1999gq,Rischke:2000qz}.

Treating the dense quark matter at moderate baryon density as a BCS-type
superconductor, one may next switch on the colored gauge fields perturbatively.
Within the effective Lagrangian \eqref{GL_effective_Lagr}, this amounts to
replacing the ordinary derivatives with the covariant ones,
$$
\de_{\mu}\Phi\to D_{\mu}\Phi=\partial_{\mu}\Phi-\imag g A_{\mu}^{a}
\left(\tfrac{1}{2}\lambda_{a}\Phi+\Phi\tfrac{1}{2}\tr\lambda_{a}\right),
$$
and adding the Yang--Mills kinetic term for the gluons. As a result of the
usual Higgs mechanism, both electric and magnetic gluons acquire nonzero masses
-- the Debye and the Meissner ones, respectively. At zero temperature, the
coefficients are roughly $\alpha_{e,m}\sim\mu^2/\Delta^2$ so that both Debye
and Meissner masses are found to be of order $g\mu$ (for detailed results and
their discussion see Ref. \pubcite1).

However, as pointed out by Rischke \cite{Rischke:2000qz}, the gauged
lowest-order Lagrangian \eqref{GL_effective_Lagr} does not reproduce correctly
the mass ratios of the gluons of different adjoint colors. The reason is the
restriction to operators of dimension four or less we employed to construct the
Lagrangian \eqref{GL_effective_Lagr}. For a more proper treatment, higher-order
operators like $\left|\Tr(\Phi^{\dagger}D_i\Phi)\right|^2$ have to be included,
which also contribute to the gluon masses.

\subsection{Nambu--Jona-Lasinio model}
We shall now develop the description using the elementary quark fields. Here we
come to the point of the proper choice of the four-fermion interaction. As
already mentioned above, we do not take up any of the interactions commonly
used in literature, but rather follow a~different approach. Our motivation goes
back two decades to the paper by Hansson et al. \cite{Hansson:1982dv}. These
authors investigated the possibility of the existence of the bound states of
two gluons and classified the strength of the QCD-induced force by the total
spin and color content of the gluon pair.

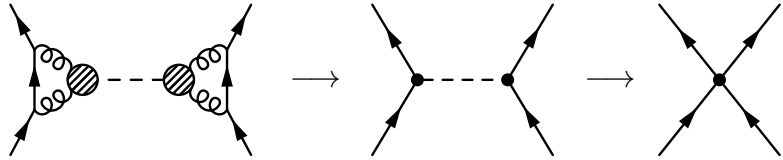
\begin{figure}
$$
\parbox{40\unitlength}{%
\begin{fmfgraph}(40,20)
\fmfset{arrow_len}{3mm}
\fmfset{curly_len}{2mm}
\fmfleftn{l}{2}
\fmfrightn{r}{2}
\fmf{fermion,tension=2}{l1,vld}
\fmf{fermion,tension=2}{vlu,l2}
\fmf{fermion}{vld,vlu}
\fmf{fermion,tension=2}{r1,vrd}
\fmf{fermion,tension=2}{vru,r2}
\fmf{fermion}{vrd,vru}
\fmf{gluon}{vld,vl,vlu}
\fmf{gluon}{vru,vr,vrd}
\fmfv{d.sh=circle,d.si=0.1w,d.fi=shaded}{vl}
\fmfv{d.sh=circle,d.si=0.1w,d.fi=shaded}{vr}
\fmf{dashes}{vr,vl}
\end{fmfgraph}}
\longrightarrow
\parbox{30\unitlength}{%
\begin{fmfgraph}(30,20)
\fmfset{arrow_len}{3mm}
\fmfleftn{l}{2}
\fmfrightn{r}{2}
\fmf{fermion}{l1,vl,l2}
\fmf{fermion}{r1,vr,r2}
\fmf{dashes}{vl,vr}
\fmfdot{vl}
\fmfdot{vr}
\end{fmfgraph}}
\longrightarrow
\parbox{20\unitlength}{%
\begin{fmfgraph}(20,20)
\fmfset{arrow_len}{3mm}
\fmfleftn{l}{2}
\fmfrightn{r}{2}
\fmf{fermion}{l1,v,l2}
\fmf{fermion}{r1,v,r2}
\fmfdot{v}
\end{fmfgraph}}
$$
\caption{Effective four-quark interaction induced by the exchange of the scalar
color-octet glueball.}
\label{Fig:glueball_induced_interaction}
\end{figure}

They discovered that apart from the colorless glueball, the most strongly bound
state is that of total spin zero which transforms as a~color octet. Such
a~state, of course, cannot exist as an excitation of the QCD vacuum. It might,
however, be a~well defined degree of freedom in the dense deconfined phase. Now
if it really exists, it certainly interacts with the quarks and its exchange
leads to the effective fermionic Lagrangian (see Fig.
\ref{Fig:glueball_induced_interaction}),
\begin{equation}
\LA=\bar\psi(\imag\slashed\de-m+\mu\gamma_0)\psi+G(\bar\psi\vek\lambda\psi)^2,
\label{NJLlagrangian}
\end{equation}
with $G>0$. (The color and spin indices are suppressed.) The proposed
interaction is attractive in the color-sextet channel and provides the basis
for the following analysis.

We use the method outlined in Section \ref{Sec:NJL_model}. Anticipating the
color-sextet condensate, we split the full Lagrangian \eqref{NJLlagrangian} in
such a~way that the free part, which determines the propagator, reads
$$
\LA_{\text{free}}=\bar\psi(\imag\slashed\partial-m+\mu\gamma_0)\psi+
\frac12\bar\psi\Delta(C\gamma_5)\tr{\bar\psi}-
\frac12\tr\psi\he\Delta(C\gamma_5)\psi.
$$
Here $\Delta$ stands for the diagonal matrix of the order parameters.

This Lagrangian is conveniently diagonalized with the help of the Nambu--Gorkov
notation,
$$
\Psi=\left(
\begin{array}{c}
\psi \\ \tr{\bar\psi}
\end{array}\right).
$$
We find, for each color $i$, two types of fermionic quasiparticles -- a
quark-like and an antiquark-like -- whose dispersion relations are
$$
E^2_{i\pm}(\vek k)=\left(\sqrt{\vek k^2+m^2}\pm\mu\right)^2+\abs{\Delta_i}^2.
$$

In the mean-field approximation the gaps $\Delta_i$ are determined by the
requirement of the cancelation of the one-loop corrections. We obtain three
separate but identical gap equations. Integrating over the frequency and
regulating the three-dimensional integral with a~cutoff $\Lambda$ they read, at
finite temperature $T$,
$$
1=\frac23G\int^{\Lambda}\frac{\dthree\vek k}{(2\pi)^3}\left(\frac1{E_+(\vek k)
}\tanh\frac{E_+(\vek k)}{2T} +\frac1{E_-(\vek k)}\tanh\frac{E_-(\vek k)
}{2T}\right).
$$

Several remarks to this result are in order. First, in its derivation we have
not been entirely self-consistent. We compared the terms of the same structure,
$\bar\psi\Delta(C\gamma_5)\tr{\bar\psi}$, in the free Lagrangian and the
one-loop correction. The presence of the chemical potential induces, however,
a~similar term $\bar\psi\Delta(C\gamma_5)\gamma_0\tr{\bar\psi}$ at one loop,
and this has been neglected. Since the full Lorentz invariance is broken by the
chemical potential, it is natural that such a~term appears. To be fully
self-consistent, we would have to include such operators into our Lagrangian
from the very beginning and solve a~coupled set of gap equations for their
coefficients. Such an analysis was done in Ref. \cite{Buballa:2001wh}.

Second, note that we derived three identical gap equations for the order
parameters $\Delta_1,\Delta_2,\Delta_3$. Since the integrands in the gap
equation are monotonic in $\Delta$, there is obviously only one nonzero
solution and thus all the gaps acquire the same value. This means that the
four-quark interaction we chose prefers the $\mathrm{SO(3)}$ symmetric phase
discussed above. This might, however, be just an artifact of the mean-field
approximation. Indeed, the separation of the three colors occurs only at the
one-loop level. The physical picture is such that the quarks of any individual
color generate a~mean field which is in turn felt only by the quarks of the
same color. It is then not surprising that all the three gaps have equal size.
At two or more loops the colors start to mix and this might lead to lifting the
degeneracy and splitting of the gaps. As shown above, if this happens the color
$\mathrm{SU(3)}$ invariance is completely broken. A~definite answer may be
given only after a~more sophisticated approximation is employed.

\section{Two-color QCD: Chiral perturbation theory}
We have already mentioned that realistic QCD calculations from first principles
are not available at moderate baryon density because of the large coupling
constant. The trouble is that neither are the lattice simulations. The reason
is that the Euclidean Dirac operator, $\mathcal
D=\gamma_{\nu}(\de_{\nu}-A_{\nu})+m-\mu\gamma_0$, is complex at nonzero baryon
chemical potential $\mu$.

This gave rise to interest in QCD-\emph{like} theories that do not have the
sign problem \cite{Kogut:1999iv}. There are two distinguished classes of such
theories -- QCD with quarks in the adjoint representation of $\mathrm{SU(3)}$
and two-color QCD \cite{Kogut:2000ek}. In the following, we shall consider the
latter case.

It turns out that the determinant of the Euclidean Dirac operator of two-color
QCD, defining the path-integral measure for the gauge bosons, is in general
just real. In order for it to be positive, there must be an even number of
quarks with the same quantum numbers \cite{Splittorff:2000mm}. Therefore, the
case of an even number of flavors is usually studied.

\subsection{Symmetry}
The key feature of the two-color QCD is the pseudoreality of the gauge group
generators, the Pauli matrices, $T_k^*=-T_2T_kT_2$. Assuming the quarks in the
fundamental (doublet) representation of the gauge $\mathrm{SU(2)}$, the
right-handed component of the Dirac spinor, $\psi_R$ (color and flavor indices
are suppressed), may be traded for the left-handed spinor
$\tilde\psi_R=\sigma_2T_2\psi_R^*$, the Pauli matrices $\sigma_k$ acting in the
Dirac space. The conjugate left-handed spinor has the same transformation
properties as $\psi_L$ and is used to replace the conventional Dirac spinor
with
$$
\Psi=\left(\begin{array}{c}
\psi_L \\
\tilde\psi_R
\end{array}\right).
$$

The Euclidean Lagrangian of massive two-color QCD at finite chemical potential
thus becomes
\begin{equation}
\LA=\imag\he\Psi\sigma_{\nu}(D_{\nu}-
\Omega_{\nu})\Psi-m\left[\tfrac12\tr\Psi\sigma_2T_2M\Psi+\text{H.c.}\right].
\label{micro_Lagrangian}
\end{equation}
Now $D_{\nu}$ is the gauge-covariant derivative and $\Omega_{\nu}$ is the
constant external field that accounts for the effects of the chemical
potential. Finally, $M$ is the block matrix in the $\Psi$ space,
$$
M=\left(\begin{array}{rc}
0 & 1\\
-1 & 0
\end{array}\right).
$$

Using the new spinor $\Psi$ it is easily seen that instead of the naively
expected chiral $\mathrm{SU(N_f)_L\times SU(N_f)_R}$ symmetry, the Lagrangian
\eqref{micro_Lagrangian} is, in the chiral limit $m=0$ and at $\Omega_{\nu}=0$,
invariant under an extended group $\mathrm{SU(2N_f)}$. At zero chemical
potential, this symmetry is broken by the standard chiral condensate down to
its $\mathrm{Sp(2N_f)}$ subgroup \cite{Kogut:2000ek}.

In the $\Psi$ notation, the standard chiral transformations correspond to
independent unitary rotations of the upper and lower components $\psi_L$ and
$\tilde\psi_R$, respectively. The new transformations in the extended group
$\mathrm{SU(2N_f)}$ mix these and thus break the baryon number. In terms of the
order parameters, these transformations rotate the chiral condensate
$\langle\bar\psi\psi\rangle$ into the diquark condensate
$\langle\psi\psi\rangle$.

It is therefore not surprising that the chemical potential term breaks the
$\mathrm{SU(2N_f)}$ down to the conventional chiral subgroup
$\mathrm{SU(N_f)_L\times SU(N_f)_R\times U(1)_B}$. The reason is that it lifts
the degeneracy between the particles and antiparticles, and the transformations
breaking the baryon number $\mathrm{U(1)_B}$ therefore no longer leave the
Lagrangian invariant.

Unlike the case of the real, three-color QCD, the two-color QCD has the
remarkable property that two quarks may form a~color-singlet state. This is
again connected to the pseudoreality of the fundamental representation of the
gauge group. It follows that the ordered phase with quarks Cooper-paired should
not be called color-superconducting, but rather just superfluid.

On the technical level, this fact has the far-reaching consequence that the
superfluidity of two-color QCD\footnote{One should carefully distinguish the
Bose--Einstein condensation of Goldstone bosons with the quantum numbers of the
diquark, from the Cooper pairing of quarks near the Fermi sea. Both effects
result in the baryon number superfluidity, but while the former occurs in the
confined phase, the latter arises from the pairing interaction between
deconfined quarks. The nice feature of two-color QCD is that the diquark
condensate may be used as an order parameter in both the confined and the
deconfined regime.} may be investigated within the framework of the chiral
perturbation theory. The effective Lagrangian is constructed on the coset space
$\mathrm{SU(2N_f)/Sp(2N_f)}$. This effective theory has been investigated to
great detail, including both the loop \cite{Splittorff:2001fy} and finite
temperature \cite{Splittorff:2002xn} effects.

The Goldstone bosons are, as usual, generated from the ground state by
space\-time-dependent symmetry transformations. In this case, they are
parametrized by an antisymmetric unimodular unitary matrix $\Sigma$. The
leading-order low-energy effective Lagrangian reads
\begin{equation}
\LA_{\text{eff}}=\frac{F^2}2\Tr(\nabla_{\nu}\Sigma\nabla_{\nu}\he\Sigma)-
G\re\Tr(J\Sigma),
\label{eff_LagrangianCHPT}
\end{equation}
where the $\nabla$'s denote the covariant derivatives,
$$
\nabla_{\nu}\Sigma=\de_{\nu}\Sigma-(\Omega_{\nu}\Sigma+\Sigma\tr\Omega_{\nu}),\quad
\nabla_{\nu}\he\Sigma=\de_{\nu}\he\Sigma+(\he\Sigma\Omega_{\nu}+\tr\Omega_{\nu}\he\Sigma),
$$
and $J$ is a~source field for $\Sigma$. When the quark mass is included, the
Goldstone bosons acquire nonzero mass $m_{\pi}$ which is related to the quark
mass $m$ by the Gell-Mann--Oakes--Renner relation
$$
mG=F^2m_{\pi}^2.
$$

In the following we shall concentrate on the simplest case $\mathrm{N_f}=2$.
Here one can take advantage of the Lie algebra isomorphisms
$\mathrm{SU(4)\simeq SO(6)}$ and $\mathrm{Sp(4)\simeq SO(5)}$. We shall argue
that it is more convenient to describe the low-energy effective theory on the
coset space $\mathrm{SO(6)/SO(5)}$.

\subsection{Coset space}
The coset $\mathrm{SU(4)/Sp(4)}$ is parametrized by the antisymmetric
unimodular unitary matrix $\Sigma$, while the coset $\mathrm{SO(6)/SO(5)}$
corresponds to the unit sphere $S^5$ i.e., it is described by a~unit vector
$\vek n$ in the six-dimensional Euclidean space. The mapping between these two
formalisms is provided by the relation
$$
\Sigma=n_i\Sigma_i,
$$
where $\Sigma_i$ are a~set of six conveniently chosen matrices, satisfying the
identity $\he\Sigma_i\Sigma_j+\he\Sigma_j\Sigma_i=2\delta_{ij}$. One particular
realization of the basis matrices is given by
\begin{equation*}
\begin{split}
\Sigma_1=\left(
\begin{array}{cr}
0 & -1\\
1 & 0
\end{array}\right),\quad
\Sigma_2=\left(
\begin{array}{cc}
\tau_2 & 0\\
0 & \tau_2
\end{array}\right),\quad
\Sigma_3=\left(
\begin{array}{cc}
0 & \imag\tau_1\\
-\imag\tau_1 & 0
\end{array}\right),\\
\Sigma_4=\left(
\begin{array}{cc}
\imag\tau_2 & 0\\
0 & -\imag\tau_2
\end{array}\right),\quad
\Sigma_5=\left(
\begin{array}{cc}
0 & \imag\tau_2\\
\imag\tau_2 & 0
\end{array}\right),\quad
\Sigma_6=\left(
\begin{array}{cc}
0 & \imag\tau_3\\
-\imag\tau_3 & 0
\end{array}\right).
\end{split}
\end{equation*}

This particular choice of the basis is not accidental. The first three matrices
have been used in literature to denote the chiral, diquark, and isospin
condensate, respectively \cite{Kogut:2000ek,Splittorff:2000mm}. The physical
nature of the individual matrices is made more transparent by assigning to them
quark bilinears,
$$
\Sigma\to\tfrac12\Psi^T\sigma_2T_2\Sigma\Psi+\text{H.c.},
$$
that provide the interpolating fields for the Goldstone bosons correspondingly.

Concretely, we find that $\Sigma_2$ and $\Sigma_4$ are real and imaginary parts
of an isospin singlet with baryon number $+1$, the diquark. Further,
$\Sigma_3,\Sigma_5,\Sigma_6$ form an isospin triplet with no baryon charge --
the pion. Finally, $\Sigma_1$ corresponds to the isospin singlet with no baryon
charge i.e., the $\sigma$ field,
\begin{gather*}
\Sigma_2\to-\tfrac12\psi^TC\gamma_5T_2\tau_2\psi+\text{H.c.},\quad
\Sigma_4\to-\tfrac12\imag\psi^TC\gamma_5T_2\tau_2\psi+\text{H.c.},\\
\Sigma_3\to-\imag\bar\psi\tau_1\gamma_5\psi,\quad
\Sigma_5\to\imag\bar\psi\tau_2\gamma_5\psi,\quad
\Sigma_6\to-\imag\bar\psi\tau_3\gamma_5\psi,\\
\Sigma_1\to\bar\psi\psi.
\end{gather*}

\subsection{Effective Lagrangian}
We shall now rewrite the effective Lagrangian in terms of the unit vector $\vek
n$. The baryon number chemical potential $\mu$ is incorporated in terms of the
external field $\Omega_{\nu}=\delta_{\nu0}\mu B$, where the baryon number
generator is represented by the block matrix
$$
B=\frac12\left(
\begin{array}{cr}
1 & 0\\
0 & -1
\end{array}\right).
$$

Adjusting the source $J$ to reproduce the quark mass effect, the leading-order
Lagrangian \eqref{eff_LagrangianCHPT} becomes
\begin{equation}
\LA_{\text{eff}}=2F^2(\de_{\nu}\vek n)^2+4\imag
F^2\mu(n_2\de_0n_4-n_4\de_0n_2)-2F^2\mu^2(n_2^2+n_4^2)-4F^2m_{\pi}^2n_1.
\label{bilinear_effective_Lagrangian}
\end{equation}

To determine the spectrum of the theory for a~particular value of the chemical
potential, one has to find the ground state by minimizing the static part of
the Lagrangian, and then expand the Lagrangian about the minimum to second
order in the fields.

\subsubsection{Normal phase}
For $\mu<m_{\pi}$ the static Lagrangian is minimized by the conventional chiral
condensate i.e., $\vek n=(1,0,0,0,0,0)$. The five independent degrees of
freedom may be identified with $n_2,\dotsc,n_6$, and the resulting dispersion
relations are
\begin{align*}
E(\vek k)&=\sqrt{\vek k^2+m_{\pi}^2}&&\text{pion triplet $n_3,n_5,n_6$},\\
E(\vek k)&=\sqrt{\vek k^2+m_{\pi}^2}-\mu&&\text{diquark $n_2+\imag n_4$},\\
E(\vek k)&=\sqrt{\vek k^2+m_{\pi}^2}+\mu&&\text{antidiquark $n_2-\imag n_4$}.
\end{align*}

This result is exactly what we would expect. The pion triplet carries no baryon
charge so its dispersion relation is not affected at all by the chemical
potential. The dispersions of the diquark and antidiquark are split and the gap
of the diquark is getting smaller until it eventually vanishes at
$\mu=m_{\pi}$. At this point the Bose--Einstein condensation sets, breaking the
baryon number spontaneously. The diquark is the corresponding Goldstone boson.

\subsubsection{Bose--Einstein condensation phase}
When $\mu>m_{\pi}$, the vacuum condensate is given by $\vek
n=(\cos\alpha,\sin\alpha,0,0,0,0)$, where $\cos\alpha=m_{\pi}^2/\mu^2$. In the
excitation spectrum we again find the pion triplet, but now with the dispersion
$E(\vek k)=\sqrt{\vek k^2+\mu^2}$. Finally, there are two excitations, the
mixtures of the (anti)diquark and $\sigma$, whose dispersion relations are
$$
E^2_{\pm}(\vek k)=\vek
k^2+\frac{\mu^2}2(1+3\cos^2\alpha)\pm\frac{\mu}2\sqrt{\mu^2(1+3\cos^2\alpha)^2+16\vek
k^2\cos^2\alpha}.
$$
Note that the gap of the `$-$' solution vanishes so that this is the Goldstone
boson of the spontaneously broken symmetry. In accordance with the general
discussion in Chapter \ref{Chap:GBcounting}, its dispersion relation is linear
at low momentum.

Our calculations confirm the results achieved previously in literature
\cite{Kogut:2000ek,Splittorff:2000mm}. The notable advantage of the
$\mathrm{SO(6)/SO(5)}$ formalism presented here is that it allows
a~straightforward physical interpretation of the various modes, being the
linear combinations of $n_1,\dotsc,n_6$ whose quantum numbers are well known.

In particular, it turns out that the quantum numbers of the Goldstone boson in
the Bose--Einstein condensed phase change as the chemical potential increases.
Just at the phase transition point, it is the diquark, matching continuously
the diquark mode in the normal phase. On the other hand, in the extreme limit
$\mu\gg m_{\pi}$, it is just $n_4$, the imaginary part of the diquark. It is
now the linear combination of the diquark and the antidiquark and thus carries
no definite baryon number. This is, of course, hardly surprising since the
baryon number is spontaneously broken and hence is not a~good quantum number
anymore.


\chapter{Conclusions}
In the preceding three chapters the results achieved during the PhD study have
been presented. Full details of the calculations may be found in the research
papers [I--IV] that are attached at the end of this thesis. Here we give
a~short summary and outline the prospects for future work.

In Chapter \ref{Chap:GBcounting} we investigated the effects of finite chemical
potential on the pattern of symmetry breaking in a~Lorentz-invariant field
theory. With the help of the Goldstone commutator we suggested a~connection
between the vacuum densities of non-Abelian charges and the counting of the
Goldstone bosons. In the framework of the linear sigma model, we were able to
formulate, and prove, an exact counting rule.

It should be stressed, however, that we stayed all the time at the tree level.
It would be desirable to investigate whether all our conclusions survive when
loop corrections are taken into account. In particular, we expect that the
leading power-like behavior of the Goldstone boson dispersion relations does
not change, up to a~possible multiplicative factor, so that the Nielsen--Chadha
counting rule is saturated.

On the other hand, we reported that right at the phase transition the phase
velocity of the linear Goldstones vanishes, changing their type from I~to II.
We also emphasized that this is the only generic case where the Nielsen--Chadha
inequality is not saturated. Since quantum corrections are expected to be
important in the vicinity of the phase transition, this result calls for
verification at one loop.

Moreover, the effect of the quartic interaction comes into play only after the
quantum corrections are included since at the tree level, the $\lambda$ term
merely serves to stabilize the static Lagrangian. Finally, due to the nonlinear
nature of the Goldstone dispersion relations (even those of type I, because of
higher orders in the power expansion of the energy), these are kinematically
allowed to decay. It is again a~matter of the one-loop effective action to
determine the corresponding decay rates. We hope that all these issues will be
clarified soon by the one-loop calculations currently being done.

Chapter \ref{Chap:EWSB} was devoted to dynamical generation of fermion masses.
We showed that a~sufficiently strong Yukawa interaction with a~complex scalar
field may result in spontaneous breaking of the chiral symmetry. This general
mechanism may find a~particular application to the standard model of
electroweak interactions -- breaking of the chiral symmetry induces breaking of
the electroweak gauge invariance -- and thus provide an alternative to the
conventional Higgs mechanism. The extension of the present Abelian model to the
electroweak $\mathrm{SU(2)_L\times U(1)_Y}$ symmetry was sketched.

However, for sake of numerical computations, we used quite crude
simplifications. We neglected all quantum corrections but the symmetry-breaking
ones to the propagators. In particular, we neglected all vertex corrections.
Such an approximation is not really consistent with the assumed symmetry i.e.,
the Ward identities, because the broken symmetry implies that the Yukawa
interaction vertex has a~pole due to the Goldstone boson. It would be perhaps
more appropriate, for instance, to generate the Schwinger--Dyson equations from
a~symmetric effective action for the full propagators, by the method of
Cornwall, Jackiw, and Tomboulis \cite{Cornwall:1974vz}.

Our future program is first to gauge the simple Abelian model presented here.
As an exercise we plan to work out signatures of the model that distinguish it
from the Higgs mechanism. The last step is  to promote the idea to the
electroweak symmetry breaking. Then we shall, of course, have to deal with the
challenges of the phenomenological restrictions. In order to make quantitative
predictions to be compared with experimental data, the approximation used here
will have to be improved a~lot. Even though this seems to be far ahead, we
believe that the mechanism we propose may provide a~viable alternative to the
Higgs mechanism.

The last topic of this thesis, the phase diagram of quantum chromodynamics, is
discussed in Chapter \ref{Chap:QCD}. We first suggest an unconventional pairing
of quarks of a~single flavor in the color-symmetric channel. Since the total
spin of such pairs is zero, they might provide a~rival to the
color-antisymmetric spin-one pairing pursued in literature. An evidence is
provided that the pairing in the color-sextet channel may arise from the
exchange of a~color-octet scalar field, a~bound state of two gluons.

The second part of Chapter \ref{Chap:QCD} is devoted to the two-color QCD. We
propose an alternative low-energy description of the two-color QCD with two
quarks flavors, based on the $\mathrm{SO(6)/SO(5)}$ coset space. We work out in
detail the correspondence with the $\mathrm{SU(4)/Sp(4)}$ formalism used in
literature and verify the results obtained by other authors.


\chapter*{List of publications}
\label{Chap:listofpubs}
\markboth{List of publications}{List of publications}
\addcontentsline{toc}{chapter}{List of publications}
\providecommand{\href}[2]{#2}

\section*{Research papers}
\begingroup\raggedright
\begin{listofpublications}{VIII}{0}
\bibitem{Brauner:2003pj}
T.~Brauner, J.~{Ho\v{s}ek}, and R.~{S\'ykora}, {\it Color superconductor with a
  color-sextet condensate},  {\em Phys. Rev.} {\bf D68} (2003) 094004,
  [\href{http://xxx.lanl.gov/abs/hep-ph/0303230}{{\tt hep-ph/0303230}}].

\bibitem{Brauner:2005hw}
T.~Brauner and J.~{Ho\v{s}ek}, {\it Dynamical fermion mass generation by a
  strong {Yukawa} interaction},  {\em Phys. Rev.} {\bf D72} (2005) 045007,
  [\href{http://xxx.lanl.gov/abs/hep-ph/0505231}{{\tt hep-ph/0505231}}].

\bibitem{Brauner:2005di}
T.~Brauner, {\it Goldstone boson counting in linear sigma models with chemical
  potential},  {\em Phys. Rev.} {\bf D72} (2005) 076002,
  [\href{http://xxx.lanl.gov/abs/hep-ph/0508011}{{\tt hep-ph/0508011}}].

\bibitem{Brauner:2006dv}
T.~Brauner, {\it On the chiral perturbation theory for two-flavor two-color
  {QCD} at finite chemical potential},  {\em Mod. Phys. Lett.} {\bf A21} (2006)
  559--569, [\href{http://xxx.lanl.gov/abs/hep-ph/0601010}{{\tt
  hep-ph/0601010}}].
\end{listofpublications}
\endgroup

\section*{Preprints}
\begingroup\raggedright
\begin{listofpublications}{VIII}{4}
\bibitem{Brauner:2004kg}
T.~Brauner and J.~{Ho\v{s}ek}, {\it A model of flavors},
  \href{http://xxx.lanl.gov/abs/hep-ph/0407339}{{\tt hep-ph/0407339}}.
\end{listofpublications}
\endgroup

\section*{Conference proceedings}
\begingroup\raggedright
\begin{listofpublications}{VIII}{5}
\bibitem{Brauner:2003br}
T.~Brauner, {\it Color superconductivity with a color-sextet order parameter},
  in {\em WDS'03 Proceedings of Contributed Papers: Part III}
  (J.~{\v{S}afr\'ankov\'a}, ed.), pp.~544--549, Matfyzpress, 2003.

\bibitem{Brauner:2004zj}
T.~Brauner, {\it Color superconductor with a color-sextet condensate},  {\em
  Eur. Phys. J.} {\bf C33} (2004) S597--S599.

\bibitem{Brauner:2005tb}
T.~Brauner, {\it Single-flavor color superconductivity with color-sextet
  pairing},  {\em Czech. J. Phys.} {\bf 55} (2005) 9--16.

\bibitem{Brauner:2005br}
T.~Brauner and J.~{Ho\v{s}ek}, {\it Dynamical symmetry breaking and mass
  generation},  in {\em WDS'05 Proceedings of Contributed Papers: Part III}
  (J.~{\v{S}afr\'ankov\'a}, ed.), pp.~436--441, Matfyzpress, 2005.
\end{listofpublications}
\endgroup

\addcontentsline{toc}{chapter}{References}
\providecommand{\href}[2]{#2}\begingroup\raggedright\endgroup

\end{fmffile}

\begin{thebibliography}{10}

\bibitem{Guralnik:1968gh}
G.~S. Guralnik, C.~R. Hagen, and T.~W.~B. Kibble, {\it Spontaneous symmetry
  breaking and the {Goldstone} theorem},  in {\em Advances in Particle Physics}
  (R.~L. Cool and R.~E. Marshak, eds.), pp.~567--708, Wiley, 1968.

\bibitem{Burgess:1998ku}
C.~P. Burgess, {\it Goldstone and pseudo-{Goldstone} bosons in nuclear,
  particle and condensed-matter physics},  {\em Phys. Rept.} {\bf 330} (2000)
  193--261, [\href{http://xxx.lanl.gov/abs/hep-th/9808176}{{\tt
  hep-th/9808176}}].

\bibitem{Kaplan:2005es}
D.~B. Kaplan, {\it Five lectures on effective field theory},
  \href{http://xxx.lanl.gov/abs/nucl-th/0510023}{{\tt nucl-th/0510023}}.

\bibitem{Manohar:1996cq}
A.~V. Manohar, {\it Effective field theories},  in {\em Perturbative and
  Nonperturbative Aspects of Quantum Field Theory} (H.~Latal and W.~Schweiger,
  eds.), pp.~311--362, Springer-Verlag, 1997.
\newblock \href{http://xxx.lanl.gov/abs/hep-ph/9606222}{{\tt hep-ph/9606222}}.

\bibitem{Scherer:2002tk}
S.~Scherer, {\it Introduction to chiral perturbation theory},
  \href{http://xxx.lanl.gov/abs/hep-ph/0210398}{{\tt hep-ph/0210398}}.

\bibitem{Low:2001bw}
I.~Low and A.~V. Manohar, {\it Spontaneously broken spacetime symmetries and
  {Goldstone}'s theorem},  {\em Phys. Rev. Lett.} {\bf 88} (2002) 101602,
  [\href{http://xxx.lanl.gov/abs/hep-th/0110285}{{\tt hep-th/0110285}}].

\bibitem{O'Raifeartaigh:1998sf}
L.~O'Raifeartaigh, {\it Hidden symmetry},  in {\em Hidden symmetries and Higgs
  phenomena} (D.~Graudenz, ed.), pp.~1--12, PSI, 1998.

\bibitem{Straumann:1998yz}
N.~Straumann, {\it Historical and other remarks on hidden symmetries},  in {\em
  Hidden symmetries and Higgs phenomena} (D.~Graudenz, ed.), pp.~233--253, PSI,
  1998.
\newblock \href{http://xxx.lanl.gov/abs/hep-ph/9810302}{{\tt hep-ph/9810302}}.

\bibitem{Coleman:1966co}
S.~Coleman, {\it The invariance of the vacuum is the invariance of the world},
  {\em J. Math. Phys.} {\bf 7} (1966) 787.

\bibitem{Weinberg:1995v1}
S.~Weinberg, {\em The Quantum Theory of Fields}, vol.~1.
\newblock Cambridge University Press, Cambridge, first~ed., 1995.

\bibitem{Weinberg:1996v2}
S.~Weinberg, {\em The Quantum Theory of Fields}, vol.~2.
\newblock Cambridge University Press, Cambridge, first~ed., 1996.

\bibitem{Goldstone:1961eq}
J.~Goldstone, {\it Field theories with `superconductor' solutions},  {\em Nuovo
  Cim.} {\bf 19} (1961) 154--164.

\bibitem{Goldstone:1962es}
J.~Goldstone, A.~Salam, and S.~Weinberg, {\it Broken symmetries},  {\em Phys.
  Rev.} {\bf 127} (1962) 965--970.

\bibitem{Fetter:1971fw}
A.~L. Fetter and J.~D. Walecka, {\em Quantum theory of many-particle systems}.
\newblock International series in pure and applied physics. McGraw--Hill, New
  York, 1971.

\bibitem{Sannino:2002wp}
F.~Sannino, {\it General structure of relativistic vector condensation},  {\em
  Phys. Rev.} {\bf D67} (2003) 054006,
  [\href{http://xxx.lanl.gov/abs/hep-ph/0211367}{{\tt hep-ph/0211367}}].

\bibitem{Rajagopal:2000wf}
K.~Rajagopal and F.~Wilczek, {\it The condensed matter physics of {QCD}},  in
  {\em At the Frontier of Particle Physics: Handbook of QCD} (M.~Shifman, ed.),
  vol.~3, chapter~35, pp.~2061--2151.
\newblock World Scientific, 2001.
\newblock \href{http://xxx.lanl.gov/abs/hep-ph/0011333}{{\tt hep-ph/0011333}}.

\bibitem{Nambu:1961tp}
Y.~Nambu and G.~Jona-Lasinio, {\it Dynamical model of elementary particles
  based on an analogy with superconductivity. {I}},  {\em Phys. Rev.} {\bf 122}
  (1961) 345--358.

\bibitem{Nambu:1961fr}
Y.~Nambu and G.~Jona-Lasinio, {\it Dynamical model of elementary particles
  based on an analogy with superconductivity. {II}},  {\em Phys. Rev.} {\bf
  124} (1961) 246--254.

\bibitem{Klevansky:1992qe}
S.~P. Klevansky, {\it The {Nambu--Jona-Lasinio} model of quantum
  chromodynamics},  {\em Rev. Mod. Phys.} {\bf 64} (1992) 649--708.

\bibitem{Buballa:2003qv}
M.~Buballa, {\it {NJL} model analysis of quark matter at large density},  {\em
  Phys. Rept.} {\bf 407} (2005) 205--376,
  [\href{http://xxx.lanl.gov/abs/hep-ph/0402234}{{\tt hep-ph/0402234}}].

\bibitem{Miransky:2001tw}
V.~A. Miransky and I.~A. Shovkovy, {\it Spontaneous symmetry breaking with
  abnormal number of {Nambu--Goldstone} bosons and kaon condensate},  {\em
  Phys. Rev. Lett.} {\bf 88} (2002) 111601,
  [\href{http://xxx.lanl.gov/abs/hep-ph/0108178}{{\tt hep-ph/0108178}}].

\bibitem{Schaefer:2001bq}
T.~Schaefer, D.~T. Son, M.~A. Stephanov, D.~Toublan, and J.~J.~M. Verbaarschot,
  {\it Kaon condensation and {Goldstone}'s theorem},  {\em Phys. Lett.} {\bf
  B522} (2001) 67--75, [\href{http://xxx.lanl.gov/abs/hep-ph/0108210}{{\tt
  hep-ph/0108210}}].

\bibitem{Blaschke:2004cs}
D.~Blaschke, D.~Ebert, K.~G. Klimenko, M.~K. Volkov, and V.~L. Yudichev, {\it
  Abnormal number of {Nambu--Goldstone} bosons in the color-asymmetric {2SC}
  phase of an {NJL}-type model},  {\em Phys. Rev.} {\bf D70} (2004) 014006,
  [\href{http://xxx.lanl.gov/abs/hep-ph/0403151}{{\tt hep-ph/0403151}}].

\bibitem{Beraudo:2004zr}
A.~Beraudo, A.~De~Pace, M.~Martini, and A.~Molinari, {\it Spontaneous symmetry
  breaking and response functions},  {\em Ann. Phys.} {\bf 317} (2005)
  444--473, [\href{http://xxx.lanl.gov/abs/nucl-th/0409039}{{\tt
  nucl-th/0409039}}].

\bibitem{Ho:1998ho}
T.-L. Ho, {\it Spinor {Bose} condensates in optical traps},  {\em Phys. Rev.
  Lett.} {\bf 81} (1998) 742--745.

\bibitem{Ohmi:1998om}
T.~Ohmi and K.~Machida, {\it Bose--{Einstein} condensation with internal
  degrees of freedom in alkali atom gases},  {\em J. Phys. Soc. Jpn.} {\bf 67}
  (1998) 1822--1825.

\bibitem{Nielsen:1976hm}
H.~B. Nielsen and S.~Chadha, {\it On how to count {Goldstone} bosons},  {\em
  Nucl. Phys.} {\bf B105} (1976) 445--453.

\bibitem{Sannino:2001fd}
F.~Sannino and W.~Schaefer, {\it Relativistic massive vector condensation},
  {\em Phys. Lett.} {\bf B527} (2002) 142--148,
  [\href{http://xxx.lanl.gov/abs/hep-ph/0111098}{{\tt hep-ph/0111098}}].

\bibitem{Leutwyler:1994gf}
H.~Leutwyler, {\it Nonrelativistic effective {Lagrangians}},  {\em Phys. Rev.}
  {\bf D49} (1994) 3033--3043,
  [\href{http://xxx.lanl.gov/abs/hep-ph/9311264}{{\tt hep-ph/9311264}}].

\bibitem{Kapusta:1981aa}
J.~I. Kapusta, {\it {Bose--Einstein} condensation, spontaneous symmetry
  breaking, and gauge theories},  {\em Phys. Rev.} {\bf D24} (1981) 426--439.

\bibitem{Buballa:2005bv}
M.~Buballa and I.~A. Shovkovy, {\it A note on color neutrality in {NJL}-type
  models},  {\em Phys. Rev.} {\bf D72} (2005) 097501,
  [\href{http://xxx.lanl.gov/abs/hep-ph/0508197}{{\tt hep-ph/0508197}}].

\bibitem{Rajagopal:2005dg}
K.~Rajagopal and A.~Schmitt, {\it Stressed pairing in conventional color
  superconductors is unavoidable},  {\em Phys. Rev.} {\bf D73} (2006) 045003,
  [\href{http://xxx.lanl.gov/abs/hep-ph/0512043}{{\tt hep-ph/0512043}}].

\bibitem{Georgi:1982jb}
H.~Georgi, {\em Lie Algebras in Particle Physics}.
\newblock Frontiers in Physics. Perseus Books, Reading, Massachusetts,
  second~ed., 1999.

\bibitem{Weinberg:1979bn}
S.~Weinberg, {\it Implications of dynamical symmetry breaking: An addendum},
  {\em Phys. Rev.} {\bf D19} (1979) 1277--1280.

\bibitem{Susskind:1979ms}
L.~Susskind, {\it Dynamics of spontaneous symmetry breaking in the
  {Weinberg--Salam} theory},  {\em Phys. Rev.} {\bf D20} (1979) 2619--2625.

\bibitem{Chivukula:1996uy}
R.~S. Chivukula, {\it An introduction to dynamical electroweak symmetry
  breaking},  in {\em Advanced School on Electroweak Theory} (D.~Espriu and
  A.~Pich, eds.), pp.~77--114, World Scientific, 1998.
\newblock \href{http://xxx.lanl.gov/abs/hep-ph/9701322}{{\tt hep-ph/9701322}}.

\bibitem{Lane:2002wv}
K.~Lane, {\it Two lectures on technicolor},
  \href{http://xxx.lanl.gov/abs/hep-ph/0202255}{{\tt hep-ph/0202255}}.

\bibitem{King:1994yr}
S.~F. King, {\it Dynamical electroweak symmetry breaking},  {\em Rept. Prog.
  Phys.} {\bf 58} (1995) 263--310,
  [\href{http://xxx.lanl.gov/abs/hep-ph/9406401}{{\tt hep-ph/9406401}}].

\bibitem{Hill:2002ap}
C.~T. Hill and E.~H. Simmons, {\it Strong dynamics and electroweak symmetry
  breaking},  {\em Phys. Rept.} {\bf 381} (2003) 235--402,
  [\href{http://xxx.lanl.gov/abs/hep-ph/0203079}{{\tt hep-ph/0203079}}].

\bibitem{Hosek:1987gf}
J.~{Ho\v{s}ek}, {\it Model for the dynamical generation of lepton, quark, and
  intermediate boson masses},  {\em Phys. Rev.} {\bf D36} (1987) 2093--2101.

\bibitem{Margolis:1984cs}
B.~Margolis and R.~R. Mendel, {\it Fermion and weak boson masses in a composite
  model},  {\em Phys. Rev.} {\bf D30} (1984) 163--173.

\bibitem{Jackiw:1973tr}
R.~Jackiw and K.~Johnson, {\it Dynamical model of spontaneously broken gauge
  symmetries},  {\em Phys. Rev.} {\bf D8} (1973) 2386--2398.

\bibitem{Bailin:1984bm}
D.~Bailin and A.~Love, {\it Superfluidity and superconductivity in relativistic
  fermion systems},  {\em Phys. Rept.} {\bf 107} (1984) 325--385.

\bibitem{Rapp:1998zu}
R.~Rapp, T.~Schaefer, E.~V. Shuryak, and M.~Velkovsky, {\it Diquark {Bose}
  condensates in high density matter and instantons},  {\em Phys. Rev. Lett.}
  {\bf 81} (1998) 53--56, [\href{http://xxx.lanl.gov/abs/hep-ph/9711396}{{\tt
  hep-ph/9711396}}].

\bibitem{Alford:1998zt}
M.~G. Alford, K.~Rajagopal, and F.~Wilczek, {\it {QCD} at finite baryon
  density: Nucleon droplets and color superconductivity},  {\em Phys. Lett.}
  {\bf B422} (1998) 247--256,
  [\href{http://xxx.lanl.gov/abs/hep-ph/9711395}{{\tt hep-ph/9711395}}].

\bibitem{Alford:2001dt}
M.~G. Alford, {\it Color superconducting quark matter},  {\em Ann. Rev. Nucl.
  Part. Sci.} {\bf 51} (2001) 131--160,
  [\href{http://xxx.lanl.gov/abs/hep-ph/0102047}{{\tt hep-ph/0102047}}].

\bibitem{Rischke:2003mt}
D.~H. Rischke, {\it The quark-gluon plasma in equilibrium},  {\em Prog. Part.
  Nucl. Phys.} {\bf 52} (2004) 197--296,
  [\href{http://xxx.lanl.gov/abs/nucl-th/0305030}{{\tt nucl-th/0305030}}].

\bibitem{Shovkovy:2004me}
I.~A. Shovkovy, {\it Two lectures on color superconductivity},  {\em Found.
  Phys.} {\bf 35} (2005) 1309--1358,
  [\href{http://xxx.lanl.gov/abs/nucl-th/0410091}{{\tt nucl-th/0410091}}].

\bibitem{Schaefer:2005ff}
T.~Schaefer, {\it Phases of {QCD}},
  \href{http://xxx.lanl.gov/abs/hep-ph/0509068}{{\tt hep-ph/0509068}}.

\bibitem{Alford:2002rz}
M.~G. Alford, J.~A. Bowers, J.~M. Cheyne, and G.~A. Cowan, {\it Single color
  and single flavor color superconductivity},  {\em Phys. Rev.} {\bf D67}
  (2003) 054018, [\href{http://xxx.lanl.gov/abs/hep-ph/0210106}{{\tt
  hep-ph/0210106}}].

\bibitem{Buballa:2002wy}
M.~Buballa, J.~{Ho\v{s}ek}, and M.~Oertel, {\it Anisotropic admixture in
  color-superconducting quark matter},  {\em Phys. Rev. Lett.} {\bf 90} (2003)
  182002, [\href{http://xxx.lanl.gov/abs/hep-ph/0204275}{{\tt
  hep-ph/0204275}}].

\bibitem{Schur:1945ab}
I.~Schur, {\it Ein satz ueber quadratische formen mit komplexen koeffizienten},
   {\em Am. J. Math.} {\bf 67} (1945) 472--480. In German.

\bibitem{Giannakis:2001wz}
I.~Giannakis and H.-C. Ren, {\it The {Ginzburg}--{Landau} free energy
  functional of color superconductivity at weak coupling},  {\em Phys. Rev.}
  {\bf D65} (2002) 054017, [\href{http://xxx.lanl.gov/abs/hep-ph/0108256}{{\tt
  hep-ph/0108256}}].

\bibitem{Pisarski:1999gq}
R.~D. Pisarski, {\it Critical line for {H}-superfluidity in strange quark
  matter?},  {\em Phys. Rev.} {\bf C62} (2000) 035202,
  [\href{http://xxx.lanl.gov/abs/nucl-th/9912070}{{\tt nucl-th/9912070}}].

\bibitem{Rischke:2000qz}
D.~H. Rischke, {\it {D}ebye screening and {M}eissner effect in a two-flavor
  color superconductor},  {\em Phys. Rev.} {\bf D62} (2000) 034007,
  [\href{http://xxx.lanl.gov/abs/nucl-th/0001040}{{\tt nucl-th/0001040}}].

\bibitem{Hansson:1982dv}
T.~H. Hansson, K.~Johnson, and C.~Peterson, {\it The {QCD} vacuum as a glueball
  condensate},  {\em Phys. Rev.} {\bf D26} (1982) 2069--2085.

\bibitem{Buballa:2001wh}
M.~Buballa, J.~{Ho\v{s}ek}, and M.~Oertel, {\it Self-consistent parametrization
  of the two-flavor isotropic color-superconducting ground state},  {\em Phys.
  Rev.} {\bf D65} (2002) 014018,
  [\href{http://xxx.lanl.gov/abs/hep-ph/0105079}{{\tt hep-ph/0105079}}].

\bibitem{Kogut:1999iv}
J.~B. Kogut, M.~A. Stephanov, and D.~Toublan, {\it On two-color {QCD} with
  baryon chemical potential},  {\em Phys. Lett.} {\bf B464} (1999) 183--191,
  [\href{http://xxx.lanl.gov/abs/hep-ph/9906346}{{\tt hep-ph/9906346}}].

\bibitem{Kogut:2000ek}
J.~B. Kogut, M.~A. Stephanov, D.~Toublan, J.~J.~M. Verbaarschot, and
  A.~Zhitnitsky, {\it {QCD}-like theories at finite baryon density},  {\em
  Nucl. Phys.} {\bf B582} (2000) 477--513,
  [\href{http://xxx.lanl.gov/abs/hep-ph/0001171}{{\tt hep-ph/0001171}}].

\bibitem{Splittorff:2000mm}
K.~Splittorff, D.~T. Son, and M.~A. Stephanov, {\it {QCD}-like theories at
  finite baryon and isospin density},  {\em Phys. Rev.} {\bf D64} (2001)
  016003, [\href{http://xxx.lanl.gov/abs/hep-ph/0012274}{{\tt
  hep-ph/0012274}}].

\bibitem{Splittorff:2001fy}
K.~Splittorff, D.~Toublan, and J.~J.~M. Verbaarschot, {\it Diquark condensate
  in {QCD} with two colors at next-to-leading order},  {\em Nucl. Phys.} {\bf
  B620} (2002) 290--314, [\href{http://xxx.lanl.gov/abs/hep-ph/0108040}{{\tt
  hep-ph/0108040}}].

\bibitem{Splittorff:2002xn}
K.~Splittorff, D.~Toublan, and J.~J.~M. Verbaarschot, {\it Thermodynamics of
  chiral symmetry at low densities},  {\em Nucl. Phys.} {\bf B639} (2002)
  524--548, [\href{http://xxx.lanl.gov/abs/hep-ph/0204076}{{\tt
  hep-ph/0204076}}].

\bibitem{Cornwall:1974vz}
J.~M. Cornwall, R.~Jackiw, and E.~Tomboulis, {\it Effective action for
  composite operators},  {\em Phys. Rev.} {\bf D10} (1974) 2428--2445.

\end{thebibliography}
\end{document}